\documentstyle[preprint,eqsecnum,amssymb,aps]{revtex}

\newcommand{\N}{{\Bbb N}}                          
\newcommand{\Z}{{\Bbb Z}}                          
\newcommand{\Ztwo}{{\Bbb Z}_2}                     
\newcommand{\Zplus}{{\Bbb Z}^+}                    
\newcommand{\Cplus}{{\Bbb C}^+}                    
\newcommand{\cf}[1]{\langle #1 \rangle}            
\newcommand{\bra}[1]{\langle #1 \!\mid\!}          
\newcommand{\ket}[1]{\!\mid\! #1 \rangle}          
\newcommand{\no}[1]{: \! #1 \! :}                  
\newcommand{\im}{\mbox{Im} \,}                     
\newcommand{\onehalf}{\mbox{$\frac{1}{2}$}}        
\newcommand{\LR}{{}^L_R}                           
\newcommand{\RL}{{}^R_L}                           
\newcommand{\1}{\openone}                          
\newcommand{\Gammaplus}{{\Gamma}^+}                
\newcommand{\vJ}{\mbox{\boldmath $J$}}             
\newcommand{\vS}{\mbox{\boldmath $S$}}             
\newcommand{\vphi}{\bbox{\phi}}                    
\newcommand{\phdagger}{\mathop{\phantom{\dagger}}} 
\newcommand{\psiop}[1]{\psi^{\phdagger}_{#1}}      
\newcommand{\psidop}[1]{\psi^{\dagger}_{#1}}       
\newcommand{\Psiop}[1]{\Psi^{\phdagger}_{#1}}      
\newcommand{\Psidop}[1]{\Psi^{\dagger}_{#1}}       
\newcommand{\cop}[1]{c^{\phdagger}_{#1}}           
\newcommand{\cdop}[1]{c^{\dagger}_{#1}}            
\newcommand{\nop}[1]{n^{\phdagger}_{#1}}           
\newcommand{\bsigma}[1]{\mbox{\boldmath $\sigma$}^{\phdagger}_{#1}} 

\newcommand{\bml}{\begin{mathletters}}             
\newcommand{\eml}{\end{mathletters} \hspace{-5pt}} 

\begin{document}

\preprint{\tt cond-mat/9506099}

\title{Magnetic Impurity in a Luttinger Liquid:\\
A Conformal Field Theory Approach}

\author{Per Fr\"ojdh}

\address{Department of Physics\\ University of Washington\\
P.O. Box 351560\\ Seattle, Washington 98195-1560\\ USA}

\author{Henrik Johannesson}

\address{Institute of Theoretical Physics\\
Chalmers University of Technology and G\"oteborg University\\
S-412 96 G\"oteborg\\ Sweden}

\maketitle
\begin{abstract}
We study the low-temperature properties of a spin-\onehalf\ magnetic
impurity coupled to a one-dimensional interacting electron system.
Using the newly developed formalism by Affleck and Ludwig, with a scale
invariant boundary condition replacing the impurity, we exploit boundary
conformal field theory to deduce the impurity thermal and magnetic
response. In the case of only forward electron scattering off the
impurity, we predict the same critical scaling as for the two-channel
Kondo effect for non-interacting electrons, but with a novel Wilson ratio.
Backward electron scattering off the impurity destabilizes this behavior
and drives the system to a new fixed point. In the case of equal amplitudes
for forward- and backward scattering {\em (Kondo interaction)}, we show
that there are only two types of scaling behaviors consistent with the
symmetries of the problem: {\em either} a local Fermi liquid {\em or} a
critical theory with an anomalous specific heat. The latter case agrees
with a recent ``poor-man-scaling'' result proposed by Furusaki and Nagaosa.
\end{abstract}

\newpage

\section{Introduction}
\label{section1}

Quantum many-particle systems sometimes exhibit a
growth of an effective coupling at low energies, resulting in a
non-perturbative ground state. The maybe simplest example of this
phenomenon is the Kondo effect \cite{Hewson}, arising from the
exchange interaction between
a spin-\onehalf\ magnetic impurity and a gas of free
quasi-particles (``dressed electrons'') in an s-wave band. As the
temperature is decreased, the system crosses over from weak
electron-impurity coupling to strong
coupling, with a complete screening of the impurity spin at $T=0$.
The resulting ground state is of Fermi liquid
type, with the single quasi-particle wave functions acquiring a phase
shift (``one-channel Kondo effect'') \cite{Nozieres}.
The picture changes when electrons in degenerate orbital bands are
allowed to interact with the impurity, and the ground state is now
described by a non-Fermi liquid fixed point (``multi-channel Kondo
effect'') \cite{NB}.

What is the corresponding scenario for an {\em interacting one-dimensional
electron system} coupled to a magnetic impurity? The question may
soon become of experimental relevance, considering the rapid
progress in the fabrication and study of very narrow conduction
channels (``quantum wires''),
obtained, for example, by gating 2D electron gases in GaAs inversion
layers \cite{Timp}. A possible laboratory realization would be a
single conduction channel with a trapped atom containing two
(or several) spin levels.
(The related problem of tunneling through potential
barriers in 1D correlated systems is already being addressed by
experimentalists \cite{Milliken},
following the pioneering work of Kane and Fisher \cite{KF}.)
The question is also interesting considering recent work on exotic
superconductivity, where the analogue to the multi-channel
Kondo effect has been exploited \cite{EmeryKivelson}. Other
realizations of Kondo physics that have been proposed
include two-level tunneling in metallic glasses \cite{VZ}, and certain
heavy-fermion materials \cite{Cox}. Treating the effect of a magnetic
impurity in the presence of interacting electrons may offer
a new perspective on these intriguing connections.

More importantly, one is here faced with an archetype
problem of describing the interplay between direct fermion
correlations (from interaction and statistics) and correlations
induced via a coupling to a local quantum mechanical degree of freedom.
As is well known, the notion of free quasi-particles
{\em (low-temperature Fermi liquid)} breaks down in one dimension:
any arbitrarily small electron-electron interaction wipes out the
single-particle poles of the electron propagator, leaving behind
only collective charge- and spin-density excitations. In the limit
of weak electron-electron interaction, these excitations are well
described  by a system of non-interacting spin-charge separated bosonic
modes {\em (Luttinger liquid)} \cite{Haldane,Schulz}.
By adding a localized magnetic impurity, one confronts the problem
of how to incorporate its coupling to {\em single electrons} in
the description of the bosonic collective degrees of freedom.

A first attack on the problem was launched by Lee and Toner
\cite{Lee}, employing Abelian bosonization followed by a perturbative
scaling analysis of a resulting ``kink-gas'' action. For the case of
a spin-\onehalf\ impurity, and with the electron gas away from
half-filling, it was found that the Kondo temperature $T_K$ (setting
the scale for the weak-to-strong coupling crossover) depends on the
bare Kondo coupling ${\lambda}_K$ in a power-law fashion, $T_K \sim
({\lambda}_K {\tau}_0)^{2/{\eta}}$. Here ${\tau}_0$ is a short-time
cutoff, and ${\eta}$ is the exponent characterizing the equal-time
spin-spin correlation in a Luttinger liquid.
For temperatures $T < T_K$, the physics is controlled by some
strong-coupling fixed point --- as in the ordinary Kondo problem --- not
directly accessible via this kind of analysis. In a recent work,
Furusaki and Nagaosa \cite{FN} derived a set of improved scaling equations
in the weak-coupling regime, preserving the spin SU(2) symmetry of the
problem. By tentatively extending these equations to the strong-coupling
regime, Furusaki and Nagaosa conjectured that the fixed-point Hamiltonian
consists of two semi-infinite Luttinger liquids and a completely screened
impurity (decoupled spin singlet). The low-temperature impurity
contributions to the specific heat and magnetic susceptibility were
calculated to $C_{imp} \sim T^{(1/K_{\rho})-1}$ and ${\chi}_{imp} \sim T^0$,
respectively, with $K_{\rho}$ the usual Luttinger liquid charge parameter
\cite{Haldane,Schulz}. Support for this scenario can be found in earlier
work \cite{EA} on impurity spins in antiferromagnetic spin-\onehalf\
chains (a ``stripped-down'' version of the Kondo effect in a Luttinger
liquid). In the case of an external $S=\onehalf$ impurity coupled to a
single site on the chain, the impurity was found to be completely screened,
severing the chain at the impurity site.

In this paper we explore the problem using {\em exact methods},
expanding upon results announced in \cite{F-J}. We shall begin by studying
a simplified model.
Specifically, we consider a spin-\onehalf\ impurity coupled with equal
strength to two next-nearest neighbor sites on a Hubbard chain. In the
continuum limit, and with quarter-filling of the band, this becomes a
Tomonaga-Luttinger (TL) model \cite{TomLut,SolEmery} with
{\em forward electron-impurity scattering} only:
\begin{equation}
{\cal H} = {\cal H}_{TL} + {\cal H}_{F} \, , \label{H}
\end{equation}
where
\begin{eqnarray}
{\cal H}_{TL}  =  \frac{1}{2 \pi} \int dx & \biggl\{ & v_F
\biggl[ \no{ \psidop{L,\sigma}(x) i \frac{d}{dx} \psiop{L,\sigma}(x) }
- \no{ \psidop{R,\sigma}(x) i \frac{d}{dx} \psiop{R,\sigma}(x) } \biggr]
 \nonumber \\
 & + &  \frac{g}{2} \sum_{r, s = L, R}
  \no{ \psidop{r,\sigma}(x) \psiop{r,\sigma}(x) }
\no{ \psidop{s, -\sigma}(x) \psiop{s, -\sigma}(x) } \nonumber \\
 & + &  \, g \, \no{ \psidop{R,\sigma}(x) \psiop{L,\sigma}(x)
\psidop{L,-\sigma}(x) \psiop{R,-\sigma}(x) } \biggr\} \label{TL}
\end{eqnarray}
and
\begin{equation}
{\cal H}_F = \lambda \left[ \no{ \psidop{L,\sigma}(0)
\onehalf \bsigma{\sigma \mu}  \psiop{L,\mu}(0) } \cdot \vS
+ \no{ \psidop{R,\sigma}(0) \onehalf \bsigma{\sigma \mu} \psiop{R,\mu}(0) }
\cdot \vS \right] . \label{F}
\end{equation}
Here $\psiop{L,\sigma}(x)$ and $\psiop{R,\sigma}(x)$ are the left- and
right-moving components, respectively, of the electron field
$\Psiop{\sigma}(x)$, with spin projection $\sigma = \uparrow, \downarrow$,
expanded about the Fermi momenta $\pm k_F$ in the long-wavelength limit:
\begin{equation}
\label{psileftright}
\Psiop{\sigma}(x) = e^{-ik_Fx}\psiop{L,\sigma}(x)
+ e^{ik_Fx}\psiop{R,\sigma}(x).
\end{equation}
The fields are normalized such that
\begin{equation}
\label{anticomm_psi}
\{ \psiop{r,\sigma}(x), \psidop{s,\mu}(y) \} = 2\pi
 \delta_{r s}  \delta_{\sigma \mu}  \delta(x-y)
\label{psianticomm}
\end{equation}
and summation over repeated (Greek) indices is implied.
The first term in (\ref{TL}) describes free left- and right-moving electrons,
whereas the second and the third terms describe forward and backward
electron-electron scattering, respectively.
The couplings $g \ (> 0)$ and $\lambda \ (> 0)$ depend on the microscopic
parameters of the lattice model and $v_F$ is the Fermi velocity. Normal
ordering $\no{ \ }$ is carried out w.r.t. the filled Dirac sea.

We shall treat the model in (\ref{H}) using the newly
developed conformal field theory approach to quantum impurity
problems by Affleck and Ludwig \cite{Affleck,AL1,Lrev}.
The basic idea of this method is to replace the impurity by a {\em boundary
condition}, in the spirit of Nozi\`{e}res' local Fermi-liquid
theory of the ordinary Kondo effect \cite{Nozieres}. At the
low-temperature fixed point, the long-wavelength properties are described
by a conformally invariant boundary condition and {\em boundary conformal
field theory} \cite{Cardy1,Cardy2,LC} essentially determines all universal
properties. Specifically, the theory predicts all boundary scaling
operators that govern the asymptotic auto-correlation functions in the
neighborhood of the impurity. As was realized by Nozi\`{e}res many years
ago \cite{Nozieres}, the impurity response to an external bulk field is
governed by the leading irrelevant boundary operators.
Thus, knowing these, one can directly deduce the impurity critical behavior.
The difficulty, though, is to identify the right boundary condition,
although frequently the symmetries of the problem cut down the list of
candidates
to a small number. In short,
each boundary condition is associated with a selection
rule for combining the various degrees of
freedom (such as charge and spin) {\em at the boundary}.
The problem thus reduces to identifying the right selection rule.
This can be done, according to the fusion-rule hypothesis of
Affleck and Ludwig \cite{AL1,AL2} by applying conformal field theory
{\em fusion rules} \cite{AffleckLesHouches}.

With only forward electron-impurity scattering present in (\ref{F}),
the model can easily be cast on a form where boundary conformal field
theory applies. Writing the Hamiltonian in terms of charge and spin
currents (Sugawara construction \cite{AffleckLesHouches}), the effect of the
impurity is traded for a new boundary condition in the spin sector.
At this point, however, two new elements enter the problem (as compared
to the treatment of the single-impurity Kondo effect for free electrons).
First, charge and spin excitations --- although dynamically decoupled ---
are still connected via a selection rule of the type mentioned above.
This must be carefully analyzed in the basis of states which diagonalize
the interacting Hamiltonian.
Secondly, by having both left- and right-moving electrons coupled to
the impurity, left- and right-moving spin excitations are no longer
separately conserved, only the total spin remains conserved.
In technical terms (to be made precise), the chiral spin
$SU(2) \times SU(2)$ symmetry of the critical {\em bulk} theory is not
recovered at the fixed-point value of the electron-impurity coupling
$\lambda$. This symmetry breaking introduces boundary operators with
non-integer scaling dimensions, by a mechanism similar to that
operating in the two-impurity Kondo problem \cite{AL2}.

The picture that emerges is consistent with that recently suggested by
Furusaki and Nagaosa \cite{FN} in their ``poor-man scaling'' analysis of
the problem: the system renormalizes onto a low-temperature fixed point
with the same critical exponents for impurity specific heat
and susceptibility as in the two-channel Kondo problem \cite{AL1}.
Any asymmetry in the left-right electron-impurity coupling
destabilizes this critical point, driving the system to a one-channel
(Fermi-liquid like) fixed point. An interesting feature of our solution
is that the leading-correction-to-scaling boundary operator at the
symmetric fixed point is unique, in contrast to previous treatments
of two-channel Kondo physics \cite{AL1}. Specifically, for vanishing
electron-electron interaction, this implies a unique Wilson ratio
also at low temperatures.

To make contact with possible future experiments, one necessarily
has to add electron back scattering off the impurity. This is so,
since a ``real'' (Kondo type) spin exchange
\begin{equation}
{\cal H}_{K} = \lambda \ \no{ \Psidop{\sigma}(0)
\onehalf \bsigma{\sigma \mu} \Psiop{\mu}(0) } \cdot \vS \ ,
\label{EXC}
\end{equation}
with $\Psiop{\sigma}(x)$ the electron field in (\ref{psileftright}),
decomposes into
\begin{equation}
{\cal H}_{K} = {\cal H}_{F} + {\cal H}_{B} \ ,
\end{equation}
${\cal H}_{F}$ being the forward scattering term in (\ref{F}), and
\begin{equation}
{\cal H}_B = \lambda \left[ \psidop{L,\sigma}(0)
\onehalf \bsigma{\sigma \mu}  \psiop{R,\mu}(0) \cdot \vS
+ \psidop{R,\sigma}(0) \onehalf \bsigma{\sigma \mu} \psiop{L,\mu}(0)
\cdot \vS \right] .
\label{B}
\end{equation}

The back scattering term ${\cal H}_{B}$ mixes left and right electron
fields, and thus breaks both chiral spin ($SU(2) \times
SU(2)$) {\em and} charge ($U(1) \times U(1)$)
symmetry of the bulk critical theory. This results in the appearance
of a relevant boundary operator which takes the system to a new
fixed point, describing Kondo scattering in a Luttinger liquid.
Turning off the electron-electron interaction, the system may still
be represented by a two-channel Hamiltonian, but now coupled to a
magnetic impurity in only one of the channels. This is known to give
a fixed point theory with Fermi liquid exponents (as for the ordinary
one-channel Kondo problem \cite{NB,ALPC}). To include the effect of
the electron-electron interaction is a more delicate problem: the mixing
of left- and right-moving electrons in (\ref{B}) obstructs a Sugawara
construction in a basis where the interaction remains local.

Not being able to attack the problem directly at a Hamiltonian level,
we shall make the natural
assumption that the full Kondo interaction may nonetheless be described
by a renormalized boundary condition on the critical bulk theory. This is
in accord with the expected behavior of {\em any} quantum impurity
interaction, as discussed in \cite{Lrev}. Note that the relevant operator
due to the Kondo interaction only couples to the boundary, i.e. the new
fixed point in this scheme is a new boundary fixed point with the
critical bulk theory unchanged, and that all conformally invariant boundary
conditions are scale invariant and correspond to such boundary fixed points.
By demanding that the non-interacting limit is correctly reproduced, with
analytic scaling in temperature for the impurity specific heat $C_{imp}$
and susceptibility $\chi_{imp}$, it turns out that conformal invariance
together with the symmetry of the problem restricts the possible types
of critical behavior to only two: {\em Either} the theory remains a local
Fermi liquid in the presence of electron-electron interaction, {\em or}
electron correlations drive the system to a new fixed point where
(to leading order in temperature)
\begin{equation}
C_{imp} =
c_1((1/K_{\rho})-1)^2T^{(1/K_{\rho})-1} + c_2 \, T \ \ \mbox{and} \ \
\chi_{imp} = c_3 \, T^0, \label{FNs}
\end{equation}
$K_{\rho} = (1+2g/v_F)^{-1/2}$ being the Luttinger liquid charge
parameter, and $c_{1,2,3}$ amplitudes depending on the scaling fields.
The second case (\ref{FNs}) agrees with the finding by Furusaki and Nagaosa
referred to above \cite{FN}. In the conformal field theory scheme this scaling
is implied by a novel selection rule for recombining the degrees of freedom
at the impurity site, and we will discuss its properties {\em in extenso}.

It should be noted that a faithful modeling of a magnetic impurity
must allow for the possibility that the impurity carries a net charge
giving rise to a screened local potential. In the case of a Fermi liquid
its effect can be absorbed by passing to a new electronic basis
with renormalized single-particle energies (which is the reason why potential
scattering is often neglected in the ordinary Kondo problem).
However, for a Luttinger liquid the effect of a local potential is more
dramatic, as shown in \cite{KF}. For this reason, potential scattering
must here be treated on equal footing with the spin exchange
interaction.\footnote{Although a local potential is generated by the
Kondo coupling $\lambda $ \cite{Njp}, it cannot simulate a realistic case
--- such as a magnetic ion trapped in a quantum wire ---  where one
expects $V \gg \lambda $, $V$ being the strength of the screened Coulomb
potential.}  Some attempts in this direction have recently been discussed
in \cite{FG}, but we will not address the problem here.

The rest of the paper is organized as follows. In the next section we
introduce the Hubbard chain coupled to an impurity spin, and perform a
continuum limit retaining only forward electron scattering off the impurity.
In Section~\ref{section3} we derive the finite-size energy spectrum and the
corresponding spectrum of boundary operator dimensions, employing a
particular variant of Affleck and Ludwig's fusion-rule hypothesis. This
section also contains a matching of the Luttinger liquid selection
rule for combining charge and spin excitations against that from a
{\em Bethe-Ansatz} analysis of the Hubbard model \cite{Woynarovich,FK,KY}.
Employing the results for the boundary operator spectrum, the impurity
critical behavior is identified in Section~\ref{section4} as that of
the two-channel Kondo problem. In Section~\ref{section5} we then consider
the effect of adding electron back scattering off the impurity, thus
treating the full Kondo interaction in a Luttinger liquid.
Section~\ref{section6}, finally,  summarizes our results.
 Throughout the paper we try
to provide sufficient information to make it essentially
self-contained for a reader with some acquaintance with conformal theory.

\section{The Model}
\label{section2}

\subsection{The Hubbard chain}

To make the physical picture clear we start by considering an explicit
model for interacting electrons on a one-dimensional lattice, coupled to a
single $S=\onehalf$ impurity. The Hamiltonian
\begin{equation}
 {\cal H} = {\cal H}_H + {\cal H}_I
\end{equation}
consists of a periodic Hubbard chain with nearest-neighbor hopping ($t$)
and repulsive on-site interactions ($U$)
\begin{equation}
\label{Hubbard}
 {\cal H}_H = - t \sum_n ( \cdop{n,\sigma} \cop{n+1,\sigma} +
 \cdop{n+1,\sigma} \cop{n,\sigma} )
 + U \sum_n \nop{n,\uparrow} \nop{n,\downarrow} \ , \ \ U > 0,
\end{equation}
and couplings ($J_n$) of the electron spins to the impurity
\begin{equation}
\label{HI}
 {\cal H}_I = \sum_n J_n \, \cdop{n,\sigma} \onehalf \bsigma{\sigma \mu}
 \cop{n,\mu} \cdot \vS,
\end{equation}
where we implicitly sum over repeated Greek indices as before.
The electron creation and annihilation operators satisfy canonical
anti-commutation relations
\begin{equation}
\{ \cop{m,\sigma}, \cdop{n,\mu} \} = \delta_{nm} \delta_{\sigma \mu}
\, ,
\end{equation}
and the number of electrons with spin $\sigma$ at site $n$ is given by
$\nop{n,\sigma} = \cdop{n,\sigma} \cop{n,\sigma}$ (without summation).
The electron density $n_e$ is the expectation value of
$\nop{n,\uparrow} + \nop{n,\downarrow}$.

Without interactions, (\ref{Hubbard}) yields a free electron dispersion
$\epsilon(k) = -2t \cos ak $, with $a$ the lattice constant. The Fermi
surface consists of the two points $k = \pm k_F$, $k_F = n_e\pi/2a$.
For small excitations (weak interaction) one may linearize the spectrum
around the Fermi points, which gives rise to left- and right-moving
particles with velocities $\pm v_F$, $v_F = 2at \sin (\pi n_e/2)$.
In terms of electron operators one has
\begin{equation}
\label{cleftright}
\cop{n,\sigma} = e^{-ik_Fna}\cop{L,n,\sigma}
+ e^{ik_Fna}\cop{R,n,\sigma} \, ,
\end{equation}
with $\cop{L,n,\sigma}$ and $\cop{R,n,\sigma}$ referring to the
excitations around $k = -k_F$ and $k= k_F$, respectively.
The sign of $k_F$ in (\ref{cleftright}) is a matter of convention;
we use that $\cop{n,\sigma}$ is expanded in $\{e^{ikna}\}$.
These electron operators, defined on the lattice, are replaced in the
continuum limit by the Dirac fields $\psiop{r,\sigma} (x)$:
\begin{equation}
 \cop{r,n,\sigma} \sim \sqrt{\frac{a}{2\pi}} \psiop{r,\sigma} (na),
\end{equation}
with normalization given by (\ref{psianticomm}). It is then
straightforward to verify that the free part of the Hamiltonian
(\ref{Hubbard}) equals ${\cal H}_{TL}$ in (\ref{TL}) with $g=0$.

As for the electron-electron interaction, it follows
from substituting (\ref{cleftright}) in (\ref{Hubbard}) that
\begin{eqnarray}
\label{Hubbard_LR}
 {\cal H}_{H} \Big|_{t=0} = U \sum_n
 & \biggl\{ & \frac{1}{2} \sum_{r, s = L, R}
\cdop{r,n,\sigma} \cop{r,n,\sigma}
\cdop{s, n,-\sigma} \cop{s,n,-\sigma} \nonumber \\
 & + &  \cdop{R,n\sigma} \cop{L,n\sigma}
\cdop{L,n,-\sigma} \cop{R,n,-\sigma} \nonumber \\
 & + &  \left[ e^{-i2\pi n n_e} \cdop{R,n,\uparrow} \cop{L,n,\uparrow}
\cdop{R,n,\downarrow} \cop{L,n,\downarrow}  + \mbox{H.c.} \right] \biggr\} ,
\end{eqnarray}
describing forward, backward and Umklapp electron-electron scattering,
respectively. Due to phase oscillations, however, the Umklapp term cancels
away from half-filling ($n_e \neq 1$), and in the continuum limit we
recover the interaction part of (\ref{TL}) with $g = Ua/2\pi$.
In this paper we will consider gapless excitations only, which
restricts us to $U>0$ and $n_e \neq 1$.

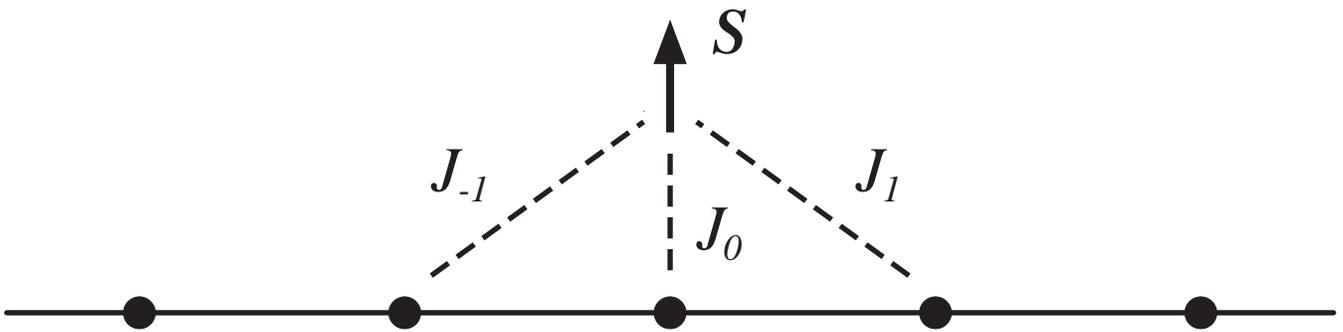
\begin{figure}
\caption{The Hubbard chain interacting with an impurity spin
$\protect{\bbox{S}}$ at a few sites. Solid lines represent electron
hopping and dashed lines spin interactions.}
\label{fig:Hubb}
\end{figure}

Finally, we rewrite the interaction with the impurity spin in (\ref{HI}).
It interacts locally with a few sites and we may concentrate on $n=0$ and
$n=\pm 1$, see Fig~\ref{fig:Hubb}. With $J_{-1} = J_1$ we recover in the
continuum limit the previous electron-impurity interactions (\ref{F}) and
(\ref{B}) with the following couplings for forward and backward scattering
against the impurity spin:
\bml
\label{lambdaFB}
\begin{eqnarray}
 \lambda_F & = & \frac{a}{2\pi} (J_0 + 2J_1), \label{lambdaF} \\
 \lambda_B & = & \frac{a}{2\pi} (J_0 + 2J_1 \cos \pi n_e ). \label{lambdaB}
\end{eqnarray}
\eml
Note that backward, but not forward, electron-impurity scattering depends on
the filling factor when the impurity is coupled to more than one
site.\footnote{There is a misprint in the corresponding formula given by
Furusaki and Nagaosa \cite{FN}. Their expression for the coupling
constant $J_B$ is valid at half-filling only, whereas the rest of their
analysis implicitly assumes other filling fractions.
This does not affect their general results, though.}
The first case of impurity
interaction addressed in this paper concerns forward scattering only, and
from this construction it is clear that one can cancel backward
scattering ($\lambda_B=0$) by coupling the impurity to the two nearest
neighboring sites at quarter filling, i.e. $J_0 = 0$ and $n_e = \onehalf$.
For other filling fractions we may also fulfill $\lambda_B=0$ by allowing
$J_0 \neq 0$. The next case of interest is the Kondo interaction,
$\lambda_F = \lambda_B \neq 0$. It corresponds to coupling the impurity
to one site only at arbitrary filling, i.e. $J_1 = 0$ in (\ref{lambdaFB}).

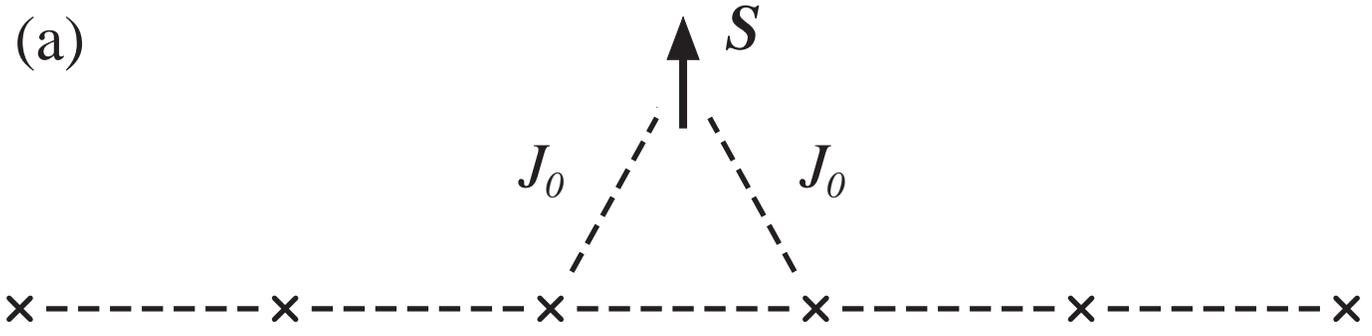
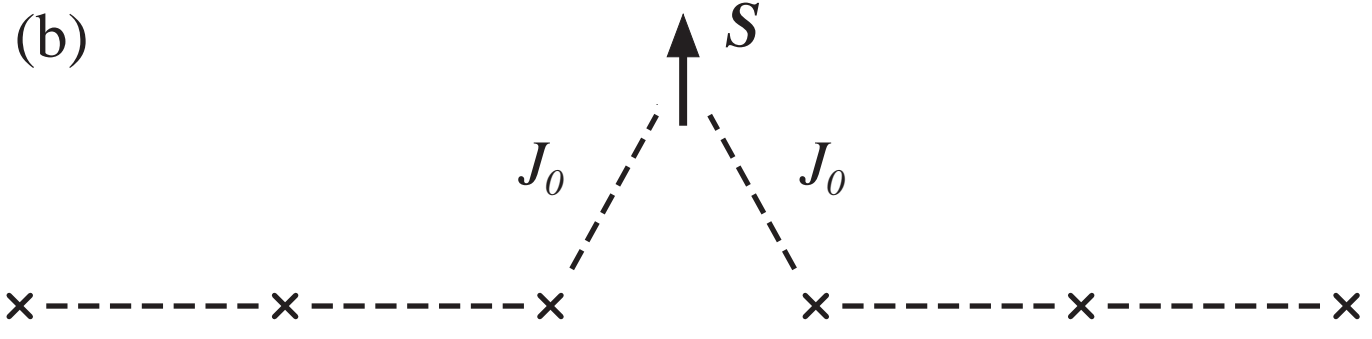
\begin{figure}
\caption{The Heisenberg antiferromagnet in 1D coupled via two neighboring
sites to an impurity $\protect{\bbox{S}}$. In (a) the chain is closed,
in (b) open, although linked via the impurity.}
\label{fig:Heis}
\end{figure}

Let us mention that another way of canceling the back-scattering term
is to couple two neighboring sites to $\vS$ at half-filling and
choose $J_0 = J_1$ (and all other $J_n = 0$). With $n_e = 1$,
Umklapp processes now come into play in (\ref{Hubbard_LR}),
causing a mass gap in the charge sector \cite{AffleckLesHouches}. The
spin sector remains massless and describes a spin-$\onehalf$
antiferromagnetic Heisenberg chain with two neighboring sites
coupled antiferromagnetically to an impurity spin (Fig.~\ref{fig:Heis}a):
\begin{equation}
{\cal H}_{spin} = J \sum_{n} \vS_{n} \cdot \vS_{n+1} +
J_0 (\vS_0 + \vS_1) \cdot \vS \, , \ \ \ \ \vS_{N} = \vS_{0} \, ,
\label{Heisenberg}
\end{equation}
with $J = 4t^2/U$.
This situation is similar to that considered by Eggert and Affleck \cite{EA},
who studied a spin chain with two open ends coupled symmetrically to a single
impurity (Fig.~\ref{fig:Heis}b). On the basis of bosonization and numerical
renormalization it was concluded that
this system is in the same universality class as the two-channel Kondo
model. The case above (Fig.~\ref{fig:Heis}a) where the spins coupled to the
impurity interact mutually has been discussed by Clarke {\em et al}.
\cite{CGS}. Also using bosonization, these authors proposed
that two-channel Kondo behavior is manifest for this case as well. Their
argument is based on a formal similarity between the bosonized
versions of the Hamiltonians for the impurity-spin chain system
and the two-channel Kondo effect, and the situation is somewhat less
clear than for the open chain. It is therefore of interest to
reconsider the problem, and we shall return to it below.

\subsection{Sugawara form}

In what follows we focus on the case away from half-filling ($n_e \neq 1$),
described by the Tomonaga-Luttinger model (\ref{TL}), together with the
forward electron-impurity interaction (\ref{F}). It is convenient to rewrite
the Hamiltonian in terms of charge and spin currents
\bml
\label{currents}
\begin{eqnarray}
J_{r}(x) & = & \ \no{ \psidop{r,\sigma}(x) \psiop{r,\sigma}(x) } \ ,
\label{U1} \\
\vJ_{r}(x) & = & \ \no{ \psidop{r,\sigma}(x) \onehalf \bsigma{\sigma \mu}
\psiop{r,\mu}(x) } \ ,
\label{SU2}
\end{eqnarray}
\eml
with $r = L$, $R$.
These obey the (level-2) $U(1)$ and level-1 $SU(2)$ affine Kac-Moody algebras
\cite{KZ}
\bml
\begin{eqnarray}
[J_{\LR}(x), J_{\LR}(y)] & = & \pm 4\pi i \, \delta'(x-y) \ ,
\label{KMU1} \\
\mbox{$[$} J^a_{\LR}(x), J^b_{\LR}(y) ] & = & i\epsilon^{abc} J^c_{\LR}
\ 2\pi \delta(x-y) \ \pm \frac{\pi i}{2} \, \delta^{ab} \delta'(x-y) \ ,
\label{KMSU2}
\end{eqnarray}
\eml
respectively, with $J^a_{r}$ the components of
$\vJ_{r} = (J^x_{r},J^y_{r},J^z_{r})$.
The normal-ordered products of the fields in (\ref{currents}) are defined
by the usual point-splitting procedure:
\begin{equation}
\no{ \psidop{r,\sigma}(x) \psiop{r,\sigma}(x)} \ \equiv
\lim_{\delta \to 0} \left[ \psidop{r,\sigma}(x+\delta)
\psiop{r,\sigma}(x) - \cf{ \psidop{r,\sigma}(x+\delta)
\psiop{r,\sigma}(x) } \right] .
\label{split}
\end{equation}
The terms of the Hamiltonian (\ref{H}) can now be identified with
combinations of the quadratic forms $\no{J_{r}(x)J_{s}(x)}$ and
$\no{\vJ_{r}(x) \cdot \vJ_{s}(x)}$ . As an example, let us consider the
case $r=s=L$. The normal ordering is again defined via point splitting,
\begin{equation}
\no{J_L(x) J_L(x)}  \  \equiv \lim_{\delta \to 0}
\left[ J_L(x+\delta) J_L(x) - \cf{J_L(x+\delta) J_L(x)} \right] ,
\end{equation}
and we need to evaluate
\begin{equation}
 J_L(x+\delta) J_L(x) = \
\no{\psidop{L,\sigma}(x+\delta) \psiop{L,\sigma}(x+\delta)} \
\no{\psidop{L,\mu}(x) \psiop{L,\mu}(x)} \ . \nonumber
\end{equation}
This can be done by using Wick's theorem and the Green's functions
\begin{equation}
 \cf{\psiop{L,\sigma}(x+\delta) \psidop{L,\mu}(x)}
 = \cf{\psidop{L,\sigma}(x+\delta) \psiop{L,\mu}(x)}
 = \frac{\delta_{\sigma \mu}}{i \delta} \ ,
\end{equation}
with the result that
\begin{eqnarray}
 J_L(x+\delta) J_L(x)
 & = &  \ \no{\psidop{L,\sigma}(x+\delta) \psiop{L,\sigma}(x+\delta)
  \psidop{L,\mu}(x) \psiop{L,\mu}(x)} \nonumber \\
 & + &  \frac{1}{i\delta}
  \no{\psiop{L,\sigma}(x+\delta) \psidop{L,\sigma}(x)}
+  \frac{1}{i\delta}
  \no{\psidop{L,\sigma}(x+\delta) \psiop{L,\sigma}(x)}
-  \frac{2}{\delta^2} \ .
\label{JJWick}
\end{eqnarray}
Hence,
\begin{equation}
\no{J_L(x) J_L(x)}  \
 =  \ \no{\psidop{L,\sigma}(x) \psiop{L,\sigma}(x)
  \psidop{L,-\sigma}(x) \psiop{L,-\sigma}(x)}
 +  2 \no{\psidop{L,\sigma}(x) i \frac{d}{dx} \psiop{L,\sigma}(x) }
\label{JJsplit}
\end{equation}
up to a total derivative from a partial integration.
The analogous procedure for the spin currents yields (using that
$\mbox{\boldmath $\sigma$}_{\sigma \mu} \cdot
\mbox{\boldmath $\sigma$}_{\nu \eta} = 2 \delta_{\sigma \eta}
 \delta_{\mu \nu} - \delta_{\sigma \mu} \delta_{\nu \eta}$)
\begin{equation}
\no{\vJ_L(x) \cdot \vJ_L(x)} \ = - \frac{3}{4} \no{\psidop{L,\sigma}(x)
\psiop{L,\sigma}(x) \psidop{L,-\sigma}(x) \psiop{L,-\sigma}(x)}
 + \frac{3}{2} \no{\psidop{L,\sigma}(x) i \frac{d}{dx} \psiop{L,\sigma}(x) }
\, ,
\label{vJvJsplit}
\end{equation}
and it is clear from (\ref{JJsplit}) and (\ref{vJvJsplit}) that the first
term of (\ref{TL}) can be written entirely in terms of currents
\begin{equation}
 \no{\psidop{L,\sigma}(x) i \frac{d}{dx} \psiop{L,\sigma}(x) } \
 = \frac{1}{4} \no{J_L(x) J_L(x)} + \frac{1}{3} \no{\vJ_L(x) \cdot
\vJ_L(x)}
\end{equation}
(up to a total derivative).
In the same manner, one straightforwardly replaces all other terms and
arrives at the Sugawara form \cite{AffleckLesHouches} of the Hamiltonian
in (\ref{H}):
\begin{eqnarray}
{\cal H}_{TL} = \frac{1}{2\pi} \int dx & \biggl\{ & \frac{v_F+g}{4}
\left[ \no{J_L(x)J_L(x)} + \no{J_R(x)J_R(x)} \right]  \nonumber \\
 & + & \frac{v_F-g}{3}  \left[ \no{\vJ_L(x) \cdot \vJ_L(x)}
+ \no{\vJ_R(x) \cdot \vJ_R(x)} \right] \nonumber \\
 & + & \, \frac{g}{2} \, \left[ J_L(x)J_R(x) - 4 \vJ_L(x) \cdot \vJ_R(x)
\right] \biggr\} \, , \label{SUGAWARAbulk}
\end{eqnarray}
and
\begin{eqnarray}
{\cal H}_{F} = \lambda \left[ \vJ_L(0) + \vJ_R(0)  \right] \cdot
\vS,
\label{SUGAWARAimp}
\end{eqnarray}
with $g = aU/2\pi$ for the Hubbard model.
Rewriting the spin-current part of the Hamiltonian on matrix form
$\no{\vJ_r(x) \cdot \vJ_r(x)} \ = \onehalf
\no{ {\cal J}_{r, \sigma \mu}(x) {\cal J}_{r, \mu \sigma}(x) }$
where ${\cal J}_{r, \sigma \mu} (x)
\equiv \ \no{\psidop{r,\sigma}(x) \psiop{r,\mu}(x)} - \onehalf
\delta_{\sigma\mu} \no{ \psidop{r,\nu}(x) \psiop{r,\nu}(x)}$
such that $\vJ_{r} (x) = \onehalf \mbox{\boldmath $\sigma$}_{\sigma \mu}
{\cal J}_{r, \sigma \mu} (x)$, one notes that the spin currents are
traceless (${\cal J}_{r, \sigma \sigma} (x) = 0$), and hence have no charge
components. Thus, spin-charge separation is manifest in (\ref{SUGAWARAbulk}).

The spin interaction term $\vJ_L(x) \cdot \vJ_R(x)$ can be shown to be
marginally irrelevant for $g > 0$ \cite{AffleckLesHouches}, and
will be dropped henceforth.\footnote{The term $\vJ_L(x) \cdot
\vJ_R(x)$ is known to give logarithmic corrections to certain
{\em bulk} correlation functions. Like any irrelevant bulk operator,
however, it cannot influence the {\em leading} boundary critical behavior
(see Sec.~\ref{section4}). Hence, we can safely drop it.}
(The spin interaction is the only term in (\ref{SUGAWARAbulk}) that
renormalizes, the remaining terms in ${\cal H}_{TL}$ being exactly marginal.)
The piece of the Hamiltonian containing charge
currents is diagonalized via the Bogoliubov transformation
\begin{equation}
J_{\LR}(x) = \mbox{cosh} \theta \, j_{\LR}(x)
- \mbox{sinh} \theta \, j_{\RL}(x) . \label{Bogo}
\end{equation}
The transformation is canonical, with the new currents $j_{\LR}
$ also satisfying the $U(1)$ Kac-Moody algebra
\begin{equation}
[j_{\LR}(x), j_{\LR}(y)] = \pm 4\pi i \, \delta'(x-y).
\label{BogoKM}
\end{equation}
Inserting (\ref{Bogo}) into (\ref{SUGAWARAbulk}), the transformation
is found to diagonalize the charge Hamiltonian when
\begin{equation}
\mbox{tanh} 2\theta = \frac{g}{v_F+g} . \label{theta}
\end{equation}
Collecting the results,
\begin{equation}
{\cal H} = {\cal H}^{*}_{TL} + {\cal H}_F
\end{equation}
with
\begin{eqnarray}
{\cal H}^{*}_{TL} = \frac{1}{2\pi} \int dx & \biggl\{ &
\frac{v_c}{4}
\left[ \no{j_L(x)j_L(x)} + \no{j_R(x)j_R(x)} \right]  \nonumber \\
 & + & \frac{v_s}{3} \left[ \no{\vJ_L(x) \cdot \vJ_L(x)}
+ \no{\vJ_R(x) \cdot \vJ_R(x)} \right] \biggr\} \label{HTLstar}
\end{eqnarray}
the resulting critical bulk Hamiltonian, and with ${\cal H}_{F}$
defined in (\ref{SUGAWARAimp}). Here
\bml
\label{velocities}
\begin{eqnarray}
v_c & = & v_F \sqrt{ 1 + \frac{2g}{v_F} } \\
v_s & = & v_F - g
\end{eqnarray}
\eml
with $v_F$ and $g$ defined above. It is easy to see that ${\cal H}^{*}_{TL}$
is invariant under independent global $U(1)$ and $SU(2)$
transformations on the left- and right-moving (chiral) fields:
\begin{equation}
\psiop{r,\sigma} \to e^{i\phi_{r}} \psiop{r,\sigma}, \ \
\psiop{r,\sigma} \to U_{r, \sigma \mu} \psiop{r,\mu}
\ ,  \ \ r = L, R\label {chiral}
\end{equation}
with $\phi_{r}$ a constant, and $U_{r,\sigma \mu}$ an element of $SU(2)$.
In fact, the Sugawara form of ${\cal H}^{*}_{TL}$ implies invariance
under the larger chiral
\begin{equation}
{\cal G} = U(1)_L \times U(1)_R \times SU(2)_{1,L}
\times SU(2)_{1,R}
\end{equation}
Kac-Moody algebra \cite{KZ} (with the subscript ``1'' denoting the level
of the $SU(2)$ algebras).
However, the chiral spin symmetry gets broken by the
impurity-electron interaction ${\cal H}_{F}$ for {\em any value of the
coupling $\lambda$ }. This is in contrast to the Kondo effect for free
electrons. In that case too there is only forward electron
scattering off the impurity, but only with one type of chiral
electrons (say, left movers). The critical bulk Kac-Moody symmetry is
$U(1) \times SU(2)_1$ (or $U(1) \times SU(2)_2 \times SU(2)_2$ in the
case of two channels), and this
symmetry is restored precisely at the strong coupling fixed
point \cite{AL1}. We return to this below.

It should be emphasized that although
the theory contains two distinct velocities, $v_c$ and $v_s$, Lorentz
invariance is manifest separately in the charge and spin sectors of
${\cal H}^{*}_{TL}$. This can be seen via bosonization
\cite{AffleckLesHouches}. The terms containing charge currents represent a
free boson theory, while the spin-current terms represent an SU(2)
$k=1$ Wess-Zumino-Witten model. Both theories are conformally invariant,
implying also Lorentz invariance. This observation will be important for
the applications to come.

\subsection{Boundary formulation}
\label{section2C}

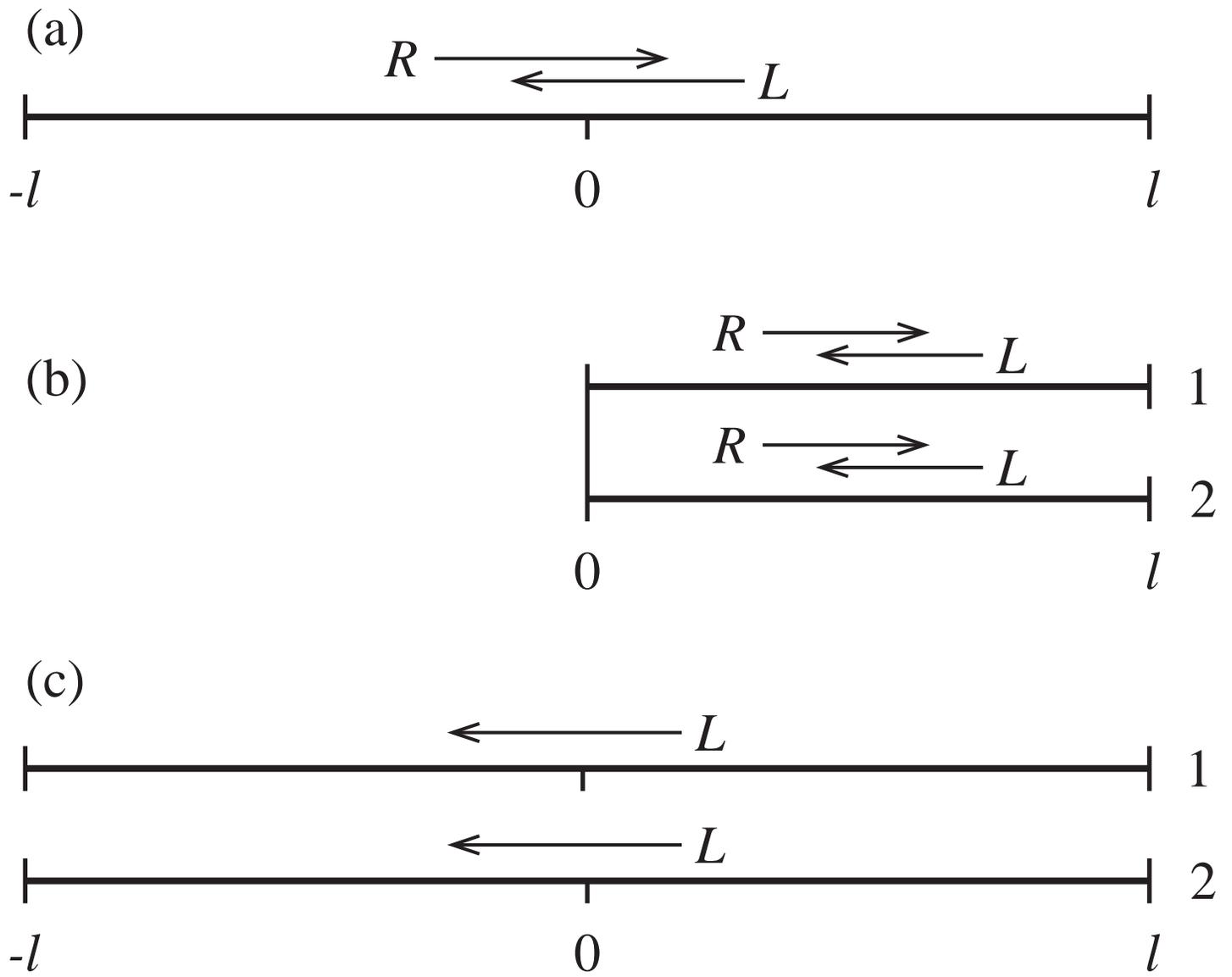
\begin{figure}
\caption{Equivalent representations of the Luttinger
liquid. The original $c=2$ system (a) with left- and right-moving excitations
on $[-\ell, \ell]$ is folded into a two-channel $c=4$ theory (b) on
$[0, \ell]$. By analytic continuation, this may be written as a chiral
$c=4$ system on $[-\ell, \ell]$ with only left-moving excitations.}
\label{fig:fold}
\end{figure}

At this point we reformulate the problem so that Cardy's
boundary conformal field theory \cite{Cardy1,Cardy2,LC} applies. The
boosted currents are defined in two-dimensional space-time $(\tau,x)$,
with the impurity sitting on the time axis. We can think of the
time axis as a boundary with periodic boundary conditions imposed on
the currents:
\begin{equation}
j_{r}(\tau, 0_+) = j_{r}(\tau, 0_-) \ , \ \ \vJ_{r}(\tau, 0_+) =
\vJ_{r}(\tau, 0_-). \label{perB}
\end{equation}
In Cardy's formalism no excitations (or two-momentum) may
flow through the boundary, and hence periodic boundary conditions
are excluded. To circumvent this restriction, we confine the system
to the interval $-\ell \leqslant x \leqslant \ell$
(taking $\ell \rightarrow \infty $ at the end), fold it in half,
double the currents, and identify the two points $x=-\ell$ and
$x=\ell$ \cite{WA}, see Figs.~\ref{fig:fold}a and~\ref{fig:fold}b.
The new currents, defined for $x \geqslant 0$ only, are related to
the old ones by
\begin{equation}
\left\{ \begin{array}{ccl} j_L^1(x) & \equiv & j_L(x) \\
   j_L^2(x) & \equiv & j_R(-x) \\ j_R^1(x) & \equiv & j_R(x) \\
   j_R^2(x) & \equiv & j_L(-x) \end{array} \right. \phantom{XXX}
\left\{ \begin{array}{ccl} \vJ_L^1(x) & \equiv & \vJ_L(x) \\
   \vJ_L^2(x) & \equiv & \vJ_R(-x) \\ \vJ_R^1(x) & \equiv & \vJ_R(x) \\
   \vJ_R^2(x) & \equiv & \vJ_L(-x) \end{array} \right.
\label{folding}
\end{equation}
and the periodic boundary conditions in (\ref{perB}) become
\begin{equation}
\left\{ \begin{array}{ccl} j_L^1(0) & = & j_R^2(0) \\
    j_L^2(0) & = & j_R^1(0) \end{array} \right. \phantom{XXXX}
\left\{ \begin{array}{ccl} \vJ_L^1(0) & = & \vJ_R^2(0) \\
    \vJ_L^2(0) & = & \vJ_R^1(0). \end{array} \right.
\label{BC}
\end{equation}
Hence, a flow of excitations across $x=0$ in the original system
corresponds to having them come in through one
channel and then reflected back through the other.

With this procedure, the critical bulk Hamiltonian is now defined on
the positive x-axis only:
\begin{eqnarray}
{\cal H}^{*}_{TL} = \frac{1}{2\pi} \sum_{i = 1,2} \int_{0}^{\ell} dx
 & \biggl\{ & \frac{v_c}{4} \left[ \no{j_L^i(x)j_L^i(x)}
+ \no{j_R^i(x)j_R^i(x)} \right] \nonumber \\
 & + & \frac{v_s}{3} \left[ \no{\vJ_L^i(x) \cdot \vJ_L^i(x)}
+ \no{\vJ_R^i(x) \cdot \vJ_R^i(x)} \right] \biggr\} \, ,\label{newSTAR}
\end{eqnarray}
with the boundary condition (\ref{BC}) imposed at $x=0$.
We also see that the electron-impurity interaction can be written as
\begin{equation}
{\cal H}_F = \lambda [ \vJ_L^1(0) +  \vJ_L^2(0) ] \cdot \vS.
\label{newFORWARD}
\end{equation}
The new Hamiltonian (\ref{newSTAR}) is invariant under the full
${\cal G} = U(1) \times U(1) \times SU(2)_{1} \times SU(2)_1$ Kac-Moody
algebra and has conformal charge $c=4$, i.e. we {\em represent} our $c=2$
theory defined on $-\ell \leqslant x \leqslant \ell$ by a $c=4$ theory on
$0 \leqslant x \leqslant \ell$.
However, we may analytically continue the left-moving currents in
(\ref{newSTAR}) to the negative x-axis\footnote{It follows from a Lagrangian
description of the Dirac fields that the left-moving currents $j_L$
and $\vJ_L$ are analytic in $z=v\tau+ix$, with $v=v_c$ and $v_s$,
respectively, whereas the right-moving currents are anti-analytic.}
 (Fig.~\ref{fig:fold}c). From (\ref{BC}) it then follows that these can be
identified with the right-moving currents on the positive axis:
\begin{equation}
\left\{ \begin{array}{ccl} j_L^1(-x) & = & j_R^2(x) \\
    j_L^2(-x) & = & j_R^1(x) \end{array} \right. \phantom{XXXX}
\left\{ \begin{array}{ccl} \vJ_L^1(-x) & = & \vJ_R^2(x) \\
    \vJ_L^2(-x) & = & \vJ_R^1(x). \end{array} \right.
\label{cont}
\end{equation}
Hence we can formulate the theory in terms of left-moving currents only.
This leads to the form of the Hamiltonian that we shall mostly use:
\begin{equation}
{\cal H} = {\cal H}_{TL}^{*} + {\cal H}_{F} \, ,
\end{equation}
where
\begin{equation}
{\cal H}^{*}_{TL} = \frac{1}{2\pi} \sum_{i = 1,2} \int_{-\ell}^{\ell} dx
\biggl\{ \frac{v_c}{4} \no{j_L^i(x)j_L^i(x)}
 + \frac{v_s}{3} \no{\vJ_L^i(x) \cdot \vJ_L^i(x)} \biggr\} \, ,
\label{Schiral}
\end{equation}
and with ${\cal H}_{F}$ given in (\ref{newFORWARD}).
The analytically continued currents satisfy periodic
boundary conditions:
\begin{equation}
j_L^i(-\ell) = j_L^i(\ell), \ \ \ \ \vJ_L^i(-\ell) =
\vJ_L^i(\ell), \ \ i= 1,2 \, , \label{pb}
\end{equation}
yielding a theory defined on a ring with circumference $2\ell$.

Before proceeding, it is instructive to look at the special
case of {\em non-interacting} ($g=0$) electrons, i.e. with $v_c =
v_s$ in (\ref{Schiral}). For this case, an alternative construction is
possible: Introducing two channels (``flavors'') of left-going fields,
\begin{equation}
\psi^1_{L,\alpha}(x) \equiv \psiop{L,\alpha} \ , \ \ \ \psi^2_{L,\alpha}
(x) \equiv -\psiop{R,\alpha}(-x) \label{simple}
\end{equation}
the free part of the bulk Hamiltonian in (\ref{TL}) attains the form
\begin{equation}
{\cal H}_0 = \frac{v_F}{2\pi} \int dx \no{\psi^{j \dagger}_{L,\sigma}(x)
i \frac{d}{dx} \psi^{j}_{L,\sigma}(x)} \label{freeH}
\end{equation}
with the electron-impurity interaction written as
\begin{equation}
{\cal H}_F = \lambda \ \no {\psi^{j \dagger}_{L,\sigma}(0) \onehalf
\bsigma{\sigma\mu} \psi^j_{L,\sigma}(0) } \cdot \vS . \label{intH}
\end{equation}
With the simple transformation in (\ref{simple}), one thus arrives at a
chiral (left-going) representation of the
{\em two-channel Kondo model} \cite{NB}. This has the Sugawara form
\cite{Affleck}
\begin{eqnarray}
{\cal H}_0 + {\cal H}_F = \frac{v_F}{2\pi} \int_{-\ell}^{\ell} dx
& \biggl\{ & \frac{1}{8} \no{J_L(x)J_L(x)} + \frac{1}{4} \no{\vJ_L(x) \cdot
\vJ_L(x)} \nonumber \\
& + & \frac{1}{4} \no{\vJ^{F}_L(x) \cdot \vJ^{F}_L(x)}
\biggr\} + \lambda \vJ_L(0) \cdot \vS \ , \label{S2channel}
\end{eqnarray}
where the currents $J_L$ (charge), $\vJ_L$ (spin), and $\vJ_L^{F}$ (flavor)
generate
the affine $U(1)$, $SU(2)_2$ and $SU(2)_2$ algebras, respectively.

The structure in (\ref{S2channel})
is not easy to obtain in the presence of electron-electron interaction.
Technically, the construction requires all excitations to have the
same velocity, which is the case only when $g=0$. The result nonetheless
suggests  that the interacting problem exhibits two-channel
Kondo behavior \cite{FN}. Since the forward electron scattering off the
impurity only affects the spin sector, the fact that the
electron-electron interaction pushes the spin and charge
excitations apart by endowing them with different velocities seems
irrelevant. However, to put the conclusion on firm ground, one must
carefully check the role of the selection rule for combining charge
and spin excitations when interactions are present.
One should here recall that although the electron-impurity
interaction is entirely in the spin sector, the charge sector may
nonetheless contribute correction-to-scaling operators, as in the
(one-channel) Kondo effect for non-interacting electrons.
A second reason for dealing with the electron-electron interaction
``head on'' is that it gives us an inroad to attack the problem of
Kondo interaction in a Luttinger liquid (by including
backward electron scattering off the impurity (see Sec.~\ref{section5})).

Returning to (\ref{Schiral}), we see that the bulk Hamiltonian ${\cal
H}^{*}_{TL}$ separately conserves the $U(1)$ and $SU(2)$ excitations
in the two channels. This simply reflects the fact that the
Kac-Moody symmetry is given by ${\cal G} = U(1) \times U(1) \times
SU(2)_1 \times SU(2)_1$, as it must. The electron-impurity
interaction, however, breaks the $SU(2)_1 \times SU(2)_1$ symmetry.
This appears similar to the effect of the Kondo interaction on free
electrons. In a left- (or right-) moving description, the Kondo term
breaks the single spin $SU(2)_1$ symmetry (or  $SU(2)_2$ in the
two-channel case) of the chiral electron Hamiltonian. However, at a
special value of the Kondo coupling, $\lambda_{Kondo} =
\lambda_{Kondo}^*$, the impurity spin can be ``absorbed'' in the
electron spin current via a canonical transformation. The impurity-electron
interaction disappears from the Hamiltonian, and the $SU(2)_1$ (or
two-channel $SU(2)_2$) symmetry is restored (now generated by the spin
current of the {\em combined} electron-impurity system). In this scheme,
$\lambda_{Kondo}^*$ {\em defines} the local strong coupling fixed
point. It is tempting to proceed in an analogous way for the present
problem, and try to absorb the impurity via the transformations
${\vJ}^i_L(x) \rightarrow {\vJ}^i_L(x) + \vS \delta (x) \ ,
i=1, 2$, judiciously choosing a special value of the
impurity-electron coupling $\lambda^*$.
However, these transformations are not canonical and couple
the two spin currents at the impurity site: the $SU(2)_1$ Kac-Moody
algebras for channel 1 and 2 are no longer independent. Thus, the
full $U(1) \times U(1) \times
SU(2)_1 \times SU(2)_1$ symmetry of ${\cal
H}^{*}_{TL}$ is {\em not} recovered at $\lambda^*$.
Now, suppose we could rewrite the spin part of ${\cal
H}^{*}_{TL}$ in terms of the {\em total} electron spin current
$\vJ (x) \equiv \vJ^{1}_L + \vJ^{2}_L$
(in addition to some auxiliary degrees of freedom).
The total current $\vJ (x)$ generates the diagonal subgroup of
$SU(2)_1 \times SU(2)_1$, and, as may be easily verified,
satisfies a level-2 $SU(2)$ Kac-Moody algebra
\begin{equation}
[J^a(x), J^b(y)] = i \epsilon^{abc} J^c(x) \ 2 \pi \delta(x-y)
 \ + \  \pi i \delta^{ab} \delta'(x-y) .
\end{equation}
The impurity may now be absorbed without problem, using the {\em
single} canonical transformation $\vJ(x) \rightarrow
\vJ(x) + \vS \delta (x) \equiv \vJ'(x)$. The combined electron-impurity
current $\vJ'(x)$ is conserved, and the $SU(2)_2$ Kac-Moody symmetry is
hence restored.

To carry out this program, one needs a dictionary to
translate from the $ U(1) \times U(1) \times
SU(2)_1 \times SU(2)_1$ formulation of the problem to another
representation in terms of $U(1) \times U(1) \times SU(2)_2 \times
{\cal G}$, where ${\cal G}$ is the symmetry group of some auxiliary
degrees of freedom. Fortunately, such a dictionary already exists in
conformal field theory (the {\em coset construction} of Goddard,
Kent, and Olive (GKO) \cite{GKO}), and provides an elegant solution to
the problem. Before exploiting it, however, we set up the formalism for
studying the finite-size spectrum of the theory.

\section{Finite-Size Spectrum and Boundary Operators}
\label{section3}

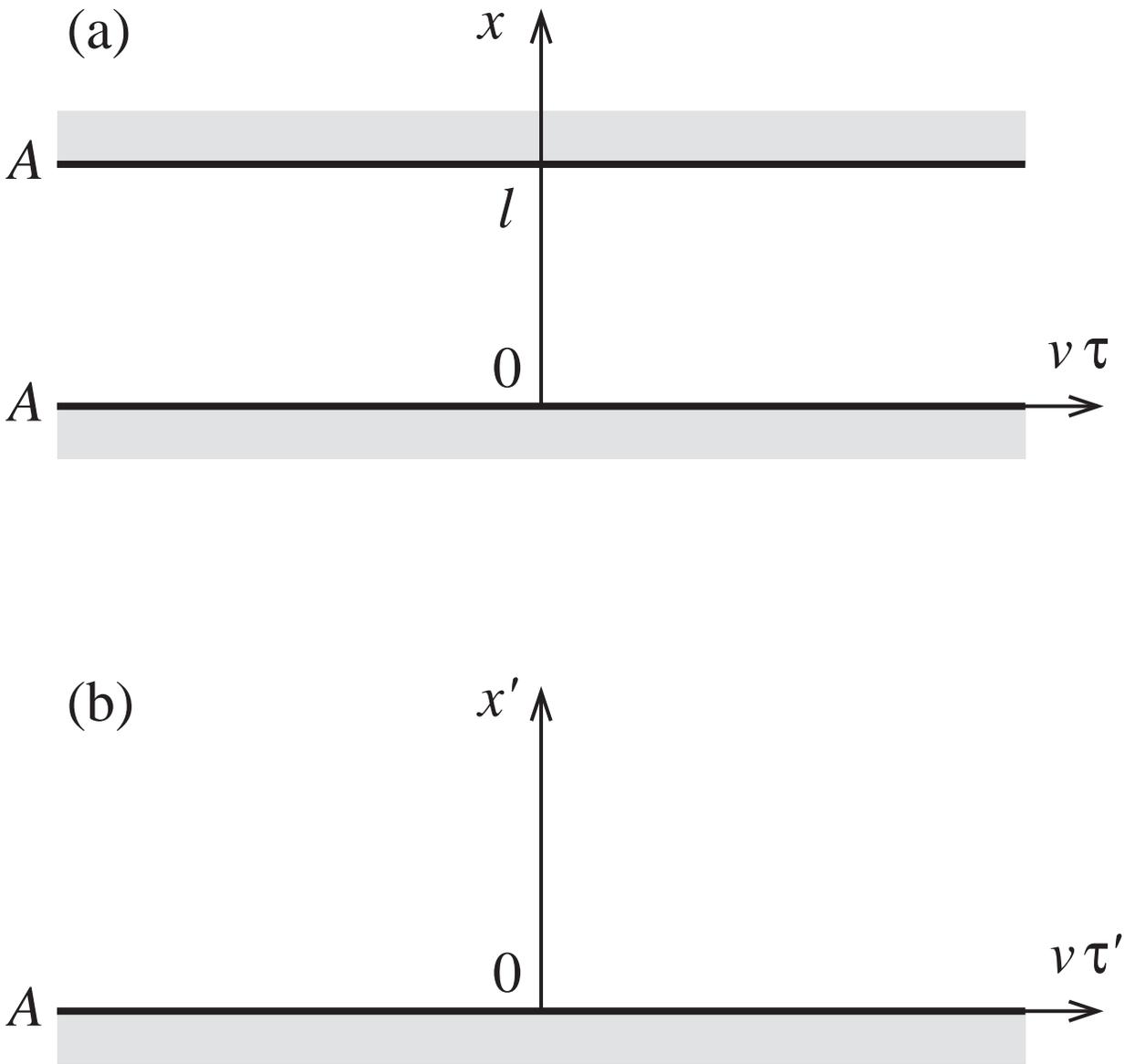
\begin{figure}
\caption{In Euclidean space-time, the finite-size theory (a)
on the strip $[0,\ell]$ with boundary
condition $A$ at both ends is conformally mapped to a semi-infinite
plane (b) with the same boundary condition $A$ applied at the boundary.}
\label{fig:strip-halfplane}
\end{figure}

The finite-size spectrum of a 1+1D scale-invariant theory provides
important information about its critical behavior. This follows from
a well-known result in conformal field theory
\cite{CardyFinite}: The energy levels in a finite geometry are
directly connected to the (boundary) scaling dimensions of operators
in the (semi-) infinite plane. More precisely, consider a conformally
invariant theory defined on the strip $\{ w=v \tau + ix \ | \
-\infty < \tau < \infty ,\  0 \leqslant x \leqslant \ell \}$ in
Fig.~\ref{fig:strip-halfplane}a,
with $\tau$ ``imaginary time'' and $x$ the space coordinate.
(The velocity of the excitations of the Hamiltonian is denoted by $v$.)
Then impose a conformally invariant boundary condition, call it $A$, at
the edges $x=0$ and $x=\ell $, and map the strip onto the semi-infinite
plane $\{z = v {\tau}' + ix' | x' \geqslant 0 \}$ in
Fig.~\ref{fig:strip-halfplane}b, using the conformal
transformation $z = \exp (\pi w/l)$ (implying boundary condition $A$
at $x' =0$). With $E^0$ the ground-state energy, one has
\begin{equation}
E = E^0 + \frac{\pi  v \Delta } {\ell}, \label{spectrum1}
\end{equation}
where $\{ E \}$ is the spectrum of excited energy levels in
$0 \leqslant x \leqslant \ell$, and $ \{ \Delta \} $ is the spectrum of
{\em boundary scaling dimensions} in the semi-infinite plane.
In a 1+1D quantum mechanical realization the boundary $x'=0$ coincides
with the time axis (${\tau}'$ also being an imaginary time), and it
follows that the boundary dimensions determine the asymptotic
auto-correlation functions. In other words, for $|{\tau}'| \gg x'$,
\begin{equation}
\cf{ {\cal{O}}({\tau}',x') {\cal{O}}(0, x') }  -
\cf{ {\cal{O}}({\tau}',x') } \cf{ {\cal{O}}(0, x') } \sim
\frac{1}{|{\tau}'|^{2\Delta}} ,
\end{equation}
with $\cal{O}$ an operator with boundary dimension $\Delta$.

For the present problem two additional features appear, not present
in the standard scenario discussed above. First, the chiral
(here, ``left-moving'') Hamiltonian in (\ref{Schiral}) represents a
full 1+1D theory on the cylinder (via the folding procedure in
(\ref{folding})): the second channel of left-moving currents simulates
the presence of right-moving currents. Therefore,
{\em bulk dimensions} appear in the finite-size scaling formula,
disguised as sums of dimensions of left-moving operators labeled by
the channel index.
Secondly, ${\cal H}^*_{TL}$ supports {\em two} kinds of
excitations, charge and spin, with distinct velocities $v_c$ and
$v_s$ (when $g \neq 0$). However, as we already noted, the
charge and spin excitations are dynamically decoupled, and
conformal invariance (including Lorentz invariance) holds separately
in the two sectors. Summing up, one expects that (\ref{spectrum1}) is
replaced by
\begin{equation}
E - E^0 = E_c + E_s - (E_c^0 + E_s^0) \, ,
\end{equation}
where
\begin{equation}
E_c - E_c^0  =  \frac{\pi v_c}{\ell}(\Delta^1_c + \Delta^2_c), \ \ \
\ \
E_s - E_s^0  =  \frac{\pi v_s}{\ell}(\Delta^1_s + \Delta^2_s) \, ,
\label{SPEC}
\end{equation}
$\{ \Delta^j_a \}$ being the
boundary dimensions in
channel
$j = 1, 2$ and sector $a=c$
{\em (charge)}, $s$ {\em
(spin)}.
(This structure of the spectrum
has been exhibited
in a {\em Bethe Ansatz} analysis of the Hubbard chain
\cite{Woynarovich}; cf. the following section.)

As we have seen in Section~\ref{section2C}, it is convenient to represent
a theory defined on a strip by a {\em chiral} theory on a cylinder.
Formally, this follows from the vanishing of the energy-momentum
tensor at the boundary \cite{Cardy1}, implying that left- and
right-moving operators coincide at the boundary.
This is precisely what we used in (\ref{cont}) when we continued
the charge and spin currents to the negative x-axis.
We know that the energy spectrum of this theory is in one-to-one
correspondence to the boundary scaling dimensions due to the boundary
conditions applied at the edges of the strip. However, these scaling
dimensions are only a subset of all possible chiral
dimensions of the bulk conformal field theory.
One is thus faced with the task to pick out those
chiral dimensions that represent the wanted boundary condition.
Formally, this may be done by connecting the boundary condition to a
{\em selection rule} that prescribes how charge and spin
excitations are combined at that boundary. Knowing the selection
rule, and allowing only boundary operators that preserve the
symmetry of the Hamiltonian, the formalism unambiguously predicts the
set of possible boundary dimensions. For a quantum impurity
problem that we are dealing with here, the more intricate task is to
identify the correct boundary condition (or, selection rule) that
represents the presence of the impurity: According to a conjecture
by Affleck and Ludwig \cite{AL1,Lrev}, any quantum impurity renormalizes,
at the fixed point, into a particular conformally invariant boundary
condition on the critical theory that carries the extended degrees of
freedom (in our case, ${\cal H}_{TL}$).
{\em One} way of identifying this boundary condition is to start with
some known, trivial
boundary condition on the critical theory, with no coupling to the
impurity. The associated selection rule simply describes the allowed
combinations of charge and spin excitations when there is no impurity
present. Now, place a
spin impurity at the boundary, and couple it to the electrons. By
redefining the spin current as that of electrons {\em and} impurity,
$\vJ(x) \rightarrow \vJ(x)
+ \vS \delta (x) \equiv \vJ'(x) $,
the electron-impurity interaction is removed at the fixed point.
The spin quantum numbers $\{j\}$ will be shifted accordingly: $\{j\}
\rightarrow \{j' \}$. The new selection rule, describing the
renormalized boundary condition, is then obtained from the old
by substituting $\{j' \}$ for $\{j\}$. The fusion-rule
hypothesis by Affleck and Ludwig \cite{AL1} suggests that the
shift of quantum numbers is precisely governed by the conformal
field theory fusion rules, in our case those of the $SU(2)_2$
Kac-Moody algebra: The set of states {\em (conformal tower)}
labeled by a quantum number $j$ is mapped onto new sets labeled by
$j'$, where
\begin{equation}
\label{SU2fusion}
j' = |j-\onehalf|, |j-\onehalf| + 1, ..., \mbox{min} \{j+\onehalf,
2-j-\onehalf\}.
\end{equation}
This is the essence of the conformal field theory approach, to be
exploited below.

In this section we study the finite-size spectrum of
${\cal H}^*_{TL}$, and derive expressions for the scaling dimensions
$\{ \Delta^j_a \}$ on a form adapted to the impurity problem. We
verify our result by matching it to that of the exact {\em Bethe-Ansatz}
analysis of the Hubbard model. Bringing the electron-impurity
interaction into play, we then use
the {\em coset construction} \cite{GKO} to make a conformal embedding of
the original $SU(2)_1 \times SU(2)_1$ spin currents
into $SU(2)_2 \times \Ztwo$. (This corresponds to writing the spin
part of ${\cal H}^*_{TL}$ as a single $SU(2)_2$
Sugawara Hamiltonian together with an Ising model.) With this
proviso, we suggest a particular application of the fusion-rule
hypothesis \cite{AL1,AL2}, and absorb the impurity spin in the
total electron spin current $\vJ(x)$, using the conformal field theory
fusion rules for the $SU(2)_2$ Kac-Moody algebra. From this, the
spectrum of boundary scaling dimensions in presence of the impurity
spin is read off.

\subsection{Finite-size spectrum of ${\cal H}^*_{TL}$}

Before applying these techniques to our Hamiltonian (\ref{Schiral}), we
need to rewrite it in Fourier space. Introducing the Fourier-transformed
currents
\begin{equation}
 \left( \begin{array}{ll}
j^i_m \\ {\vJ}^i_m
\end{array} \right)
= \frac{1}{2 \pi}
\int^{\ell}_{-\ell} dx \, e^{im\frac{\pi}{\ell}x}
\left( \begin{array}{ll}
j^i_L(x) \\ {\vJ}^i_L(x)
\end{array} \right)
\label{FTC}
\end{equation}
for $i = 1,2$ and $m \in \Z$, the mode expanded Hamiltonian takes the form
\begin{equation}
{\cal H}^*_{TL} = \sum_{i=1,2} {\cal H}^i_c + {\cal H}^i_s \, ,
\end{equation}
where
\bml
\begin{eqnarray}
{\cal H}^i_c & = & \frac{\pi v_c}{\ell} \{ \frac{1}{4}j^i_0
j^i_0 + \frac{1}{2} \sum_{m=1}^{\infty} j_{-m}^ij_m^i - \frac{1}{24} \}
\label{modeCHARGE} \\
{\cal H}^i_s & = & \frac{\pi v_s}{\ell} \{ \frac{1}{3} {\vJ
}^i_0 \cdot {\vJ }^i_0 + \frac{2}{3} \sum_{m=1}^{\infty}
{{\vJ}}^i_{-m} \cdot {\vJ}^i_m- \frac{1}{24} \} . \label{modeSPIN}
\end{eqnarray}
\eml
This result is easy to verify. Consider first the $U(1)$ currents,
with
\begin{equation}
\no{ j^i_L(x)j^i_L(x)} \ \equiv
\lim_{\delta \to 0} \left[ j^i_L(x+\delta)
j^i_L(x) - \cf{ j^i_L(x+\delta)
j^i_L(x)
 } \right] .
\label{splitU1}
\end{equation}
Noting that $j^i_L$ is a dimension-1 analytic operator with
argument $z=v_c\tau+ix$, it follows from the operator product
expansion \cite{ID} of $j^i_L$ with
itself that\footnote{The factor of 2 in the singular term comes from
the ``level'' of the U(1) currents, these being sums over $\sigma =
\uparrow, \downarrow$ fields. Cf. (\ref{JJWick}) for $J_L(x)$
that satisfies the same algebra.}
\begin{equation}
\cf{ j^i_L(x+\delta) j^i_L(x) } = \frac{2}{(i \delta)^2}
+ \frac{1}{\ell^2} O[(\frac{\delta}{\ell})^0] ,
\label{delta}
\end{equation}
where the last term is the correction due to the finite $\ell$.
The inverse of (\ref{FTC}),
\begin{equation}
\left( \begin{array}{ll}
j^i_L(x) \\ {\vJ}^i_L(x)
\end{array} \right)
= \frac{\pi}{\ell} \sum_{m = -\infty}^{\infty} e^{-im\frac{\pi}{\ell}x}
\left( \begin{array}{ll}
j^i_m \\ {\vJ}^i_m
\end{array} \right) , \label{IFTC}
\end{equation}
together with
(\ref{splitU1}) and (\ref{delta}), implies that
\begin{equation}
\frac{1}{2\pi}
\int_{-\ell}^{\ell} dx \, \no{j^i_L(x)j^i_L(x)} \
 = \lim_{\delta \to 0} \left[ \frac{\pi}{\ell} \sum_{n}
e^{-in{\frac{\pi}{\ell}}\delta} j^i_nj^i_{-n}+
\frac{2\ell}{\pi \delta^2} \right] . \label{result}
\end{equation}
Using the $U(1)$ Kac-Moody algebra (\ref{BogoKM}) in Fourier space,
\begin{equation}
[j^i_n \ , j^k_m ] = 2n \, {\delta}_{n+m, 0} \, {\delta}^{ik} \, ,
\label{U1KM}
\end{equation}
we may write
\begin{equation}
\sum_{n}
e^{-in{\frac{\pi}{\ell}}\delta} j^i_nj^i_{-n} =
j^i_0 j^i_0 + 2\sum_{n>0} \mbox{cos}(n\frac{\pi}{\ell}
\delta)j^i_{-n}j^i_n + 2 \sum_{n>0} n e^{-in \frac{\pi}{\ell} \delta },
\end{equation}
so that the last term cancels the singular part of (\ref{result}):
\begin{equation}
\frac{2\pi}{\ell} \sum_{n>0}
n e^{-in \frac{\pi}{\ell} \delta } = -\frac{2\ell}{\pi \delta^2}
- \frac{\pi}{6 \ell} + \frac{1}{\ell} O [(\frac{\delta}{\ell})^2].
\end{equation}
Hence, ${\cal H}^i_c$ (\ref{modeCHARGE}) follows from comparing
(\ref{Schiral}) with (\ref{result}). The analogous treatment of the
Fourier transformed spin currents, satisfying the $SU(2)_1$ Kac-Moody algebra
\begin{equation}
[J^{i a}_n \ , J^{k b}_m ] = i{\epsilon}^{abc}J^{i c}_{n+m}
{\delta}^{ik} + \frac{n}{2} \, \delta_{n+m, 0} \, \delta^{ab}\delta^{ik}
\, ,
\label{SU2KM}
\end{equation}
leads to (\ref{modeSPIN}). As can be seen from the derivation, the
Schwinger type terms in (\ref{modeCHARGE}) and (\ref{modeSPIN}),
including the constant $\frac{1}{24}$, are due to the regularization of
the spectrum (normal ordering of the quadratic currents). These terms
encode the conformal anomaly \cite{ID} of the theory, and we shall discuss
their role below.

Given ${\cal H}^i_c$ (\ref{modeCHARGE}) and ${\cal H}^i_s$
(\ref{modeSPIN}), it is now easy to extract the finite-size spectrum.
Let us start with the charge Hamiltonian,
${\cal H}_c = {\cal H}^1_c + {\cal H}^2_c$, and make the connection to the
original electron fields in the Tomonaga-Luttinger Hamiltonian (\ref{TL}).
By construction (cf. Eq. (\ref{folding}) and (\ref{cont})) we can
identify
\begin{equation}
j^1_m = j_{L,m},  \ \ \ \ \ \ j^2_m = j_{R,m}, \label{identify}
\end{equation}
where
\begin{equation}
j_{\LR,m} = \frac{1}{2\pi} \int_{-\ell}^{\ell} dx \, e^{\pm
im\frac{\pi}{\ell}x}j_{\LR}(x)
\end{equation}
are the Fourier transforms of the left/right-moving $U(1)$
currents introduced in (\ref{Bogo}). Thus
\begin{equation}
\left\{ \begin{array}{ccl}
j^1_m & = & \mbox{cosh} \theta J_{L,m} + \mbox{sinh} \theta J_{R,m} \\
j^2_m & = & \mbox{cosh} \theta J_{R,m} + \mbox{sinh} \theta J_{L,m},
\end{array} \right.
\label{FourierBogo}
\end{equation}
with
\begin{equation}
J_{\LR,m} = \frac{1}{2\pi} \int_{-\ell}^{\ell}
dx \, e^{\pm im \frac{\pi}{\ell} x}
\no{ \psidop{\LR,\sigma}(x) \psiop{\LR,\sigma}(x)}
\label{JFourier}
\end{equation}
the Fourier components of the original $U(1)$ currents in (\ref{U1}).
For later convenience we impose anti-periodic boundary
conditions\footnote{The condition $\psiop{r,\sigma}(-\ell)=
-\psiop{r,\sigma}(\ell)$ is consistent with periodic boundary conditions
on the currents and implies that the Fermi level lies between two energy
levels in the finite-size spectrum.} on $\psiop{r,\sigma}(x)$, which then
can be expanded as
\begin{equation}
\psiop{\LR,\sigma}(x) = \frac{\pi}{\ell} \sum_{n}
e^{\mp i(n+\onehalf)\frac{\pi}{\ell}x}
 \psiop{\LR,\sigma,n} \ .
\end{equation}
This definition implies that the momenta for
$\psiop{\LR,\sigma,n}$ are given by $k=\mp \frac{\pi}{\ell} (n + \onehalf)$,
so that the ``single-particle'' energy levels $\mp v_c k$, w.r.t.
the Fermi level $k=0$, both satisfy
\begin{equation}
 \epsilon_n = \frac{v_c \pi}{\ell} (n+\onehalf).
\end{equation}
As follows from the definition of the integrand of (\ref{JFourier}) in
terms of point splitting (\ref{split}), we need the
(finite-size) Green's function
\begin{equation}
\cf{\psidop{\LR,\sigma}(x+\delta) \psiop{\LR,\sigma}(x)} =
\frac{2}{(\pm i\delta)} + \frac{1}{\ell} O(\frac{\delta}{\ell})
\end{equation}
to obtain
\begin{equation}
J_{\LR,m} = \lim_{\delta \to 0} \left[ \frac{\pi}{\ell} \sum_{n} e^{\pm
i (n+\onehalf) \frac{\pi}{\ell} \delta} \psidop{\LR,\sigma,n}
\psiop{\LR,\sigma,n+m} \mp \frac{2\ell\delta_{m0}}{i\pi \delta} \right] \ .
\label{intermediate}
\end{equation}
It is now convenient to introduce normal ordering in
Fourier space. The filled Fermi sea occupies all levels for $n<0$,
and hence
\begin{equation}
\no{\psidop{r,\sigma,n} \psiop{r,\sigma,m}} \ = \left\{ \begin{array}{ll}
\psidop{r,\sigma,n} \psiop{r,\sigma,m} & n \neq m \
\mbox{or} \ n=m \geqslant 0 \\
-\psiop{r,\sigma,m} \psidop{r,\sigma,n} & n=m < 0, \end{array} \right.
\end{equation}
for $r=L$ or $R$. Using $\{ \psiop{r,\sigma,m}, \psidop{s,\mu,n} \} =
\frac{\ell}{\pi} \delta_{r s} \delta_{\sigma\mu} \delta_{mn}$ and
\begin{equation}
\sum_{n>0} e^{\pm i (n+\onehalf) \frac{\pi}{\ell} \delta} =
 \pm \frac{\ell}{i \pi \delta} + O(\frac{\delta}{\ell})
\end{equation}
we may finally write
\begin{equation}
J_{\LR,m} = \frac{\pi}{\ell} \sum_{n} \no{\psidop{\LR,\sigma,n}
\psiop{\LR,\sigma,n+m}} \ . \label{sumrep}
\end{equation}
The interpretation of $J_{\LR,m}$ is now straightforward: For $m=0$ it
counts the net number of left(right) moving particles w. r. t. the Fermi
sea, whereas for $m \neq 0$ it excites particles $m$ steps.

Next we introduce a set of {\em Kac-Moody primary states} $\ket{\Lambda_P}$
w.r.t. the charge and spin currents \cite{ID}, defined by
\begin{equation}
J_{\LR,m}\ket{\Lambda_P} = 0, \ \ \  \vJ_{\LR,m}\ket{\Lambda_P} = 0 \ , \ \
m > 0 ,\label{defP}
\end{equation}
from which all other states can be generated. In an occupation
number representation \cite{Llec},
\begin{equation}
\ket{\Lambda_P}  \equiv \ \ket{Q_{L\uparrow} , Q_{L\downarrow},
Q_{R\uparrow}, Q_{R
\downarrow}}
\label{PrimaryState}
\end{equation}
with the constraint
\begin{equation}
\label{PrimaryConstraints}
\mid Q_{\LR \uparrow} - Q_{\LR \downarrow} \mid  \leqslant  1 \ .
\end{equation}
Here $\ket{Q_{r \sigma}}$ denotes a non-excited state with
$Q_{r \sigma}$ the number of $r=L$, $R$ electrons, carrying
spin $\sigma$, added to the filled Fermi sea. Hence, the total charge in
channel $r$, $Q_r = Q_{r\uparrow} + Q_{r\downarrow}$, is the eigenvalue of
$J_{r,0}$ in (\ref{sumrep}). Combining (\ref{modeCHARGE}),
(\ref{FourierBogo}), and (\ref{sumrep}), we thus have
\begin{equation}
{\cal H}^i_c\ket{\Lambda_P} = \frac{\pi v_c}{\ell}
\left\{ \frac{(q^i)^2}{4} - \frac{1}{24} \right\}
\ket{\Lambda_P} \ , \ \ \ i=1,2  \label{Ceigen}
\end{equation}
where we have introduced the eigenvalues of $j^i_0$,
\begin{equation}
\label{qnumbers}
q^{{}^1_2} = Q \frac{e^\theta}{2} \pm \Delta Q \frac{e^{-\theta}}{2}
\end{equation}
labeled by the quantum numbers
\begin{equation}
\left\{ \begin{array}{rcccl}
Q & \equiv & Q_L + Q_R & = &  Q_{L\uparrow} + Q_{L\downarrow}
 + Q_{R\uparrow} + Q_{R\downarrow} \\
\Delta Q & \equiv & Q_L - Q_R & = & Q_{L\uparrow} + Q_{L\downarrow}
 - Q_{R\uparrow} - Q_{R\downarrow} \ . \end{array} \right. \label{Q}
\end{equation}
As $Q_L, Q_R \in \Z$, it follows that $Q, \Delta Q \in \Z$ with the
restriction that $Q \pm \Delta Q$ is even.

Turning to the spin Hamiltonian, ${\cal H}_s$, we proceed analogously.
Writing $\vJ^i_0 \cdot \vJ^i_0 = 3J^{iz}_0 J^{iz}_0$ in (\ref{modeSPIN}),
and identifying Fourier modes,
\begin{equation}
J^{1z}_m = J^z_{L,m} \ \ \ \ \ J^{2z}_m = J^z_{R,m} \ ,
\end{equation}
the spin Hamiltonian (\ref{modeSPIN}) and the Fourier transform of
(\ref{SU2}) imply that
\begin{equation}
{\cal H}^i_s \ket{ \Lambda_P } \ = \ \frac{\pi v_s}{\ell}
\{\frac{1}{4}(Q^i_{\uparrow} - Q^i_{\downarrow})^2 - \frac{1}{24}
\}\ket{ \Lambda_P }. \label{Seigen}
\end{equation}
(In obvious notation, $Q^1 = Q_L$ and $Q^2 = Q_R$.)
Introducing spin quantum numbers $j^i = 0, \onehalf, \ i=1,2$,
connected to the particle numbers by
\begin{equation}
\frac{4}{3} j^i(j^i+1) =  ( Q^i_{\uparrow} - Q^i_{\downarrow} )^2 \
\leqslant 1 \, , \label{connection}
\end{equation}
the eigenvalues in (\ref{Seigen}) are expressed as
\begin{equation}
E^i_s(j^i) =
\frac{\pi v_s}{\ell} \left\{ \frac{1}{3}j^i(j^i+1) - \frac{1}{24}
\right\} , \ \ \
j^i=0,\onehalf . \label{j}
\end{equation}
(This result can also be obtained directly by noting that $\vJ^i_0
\cdot \vJ^i_0$ is the Casimir invariant of $SU(2)_1$, with
eigenvalues $j^i(j^i+1)$ when acting on the primary states \cite{KZ}.)

As a consequence of $[{\cal H}_c,{\cal H}_s]=0$, the primary states
factorize in charge and spin,
\begin{equation}
\ket{ \Lambda_P } = \ \ket{ \Lambda_P }_c \ \otimes
\ket{ \Lambda_P }_s \ ,
\label{Pfactor}
\end{equation}
and as the Hamiltonians are diagonal in channels $1$ and $2$,
\bml
\begin{eqnarray}
\ket{ \Lambda_P }_s & = & \ \ket{ j^1}^{(1)} \ \otimes \ket{
j^2}^{(2)} \label{Sstate} \\
\ket{\Lambda_P }_c & = & \ \ket{q^1}^{(1)}
\ \otimes \ket{q^2}^{(2)} \ , \label{Cstate}
\end{eqnarray}
\label{SCstates}
\eml
the superscripts denoting the two channels. Note that the $q^i$'s mix
the original $L$ and $R$ channels and are defined
by (\ref{qnumbers}) in terms of the quantum numbers $Q$ and $\Delta Q$.

To obtain a state of arbitrary particle number $(Q_{L\uparrow} ,
Q_{L\downarrow}, Q_{R\uparrow}, Q_{R\downarrow})$
we undo the constraints (\ref{PrimaryConstraints}) by applying the
operators $J^{\pm}_{r,-m} =J^{x}_{r, -m} \pm J^{y}_{r, -m}$, with  $m>0$,
on $\ket{ \Lambda_P }$. These operators flip spin within the left- and
right-moving branches. The resulting states, call them $\ket{ \Lambda }$, are
still labeled by $Q_{\LR \alpha}$, which now may be any integer.
Furthermore, by acting with the operators $J_{r, -m}$ and $J^z_{r, -m}$
($m>0$) on $\ket{ \Lambda }$, we also remove the constraint that
$\ket{Q_{r \sigma}}$ is non-excited, i.e. we allow non-filled levels
below the highest occupied level: As is readily verified, the Fourier modes
$\onehalf J_{r,-m} \pm J^z_{r,-m}$ ($m>0$) create ``particle-hole
excitations'' from the states $\ket{ \Lambda }$ within each
$Q_{r\sigma}$-branch.
{\em Any} state of particle number $(Q_{L\uparrow}, Q_{L\downarrow}, Q_{R
\uparrow}, Q_{R\downarrow})$ can thus be obtained from the set of
primary states. (In the following, we label our states in the diagonal basis
of ${\cal H}$, though.) As follows from (\ref{FourierBogo}) and (\ref{defP}),
the primary states $\ket{\Lambda_P}$ are also primary w.r.t. to the
diagonalized currents $j^i_{-m}$ and $\vJ^i_{-m}$ and {\em any} other
excited state is obtained by applying these operators for $m>0$.

To extract the spectrum including the energy levels of the
{\em descendant states} just exposed, it is sufficient to note that the
operators $ \{ j^k_{-m}; m>0, k=1,2 \}$ and $ \{ \vJ^k_{-m}; m>0, k=1,2 \}$
act as ``raising operators'' w.r.t. the primary states in (\ref{Sstate})
and (\ref{Cstate}). Explicitly, from (\ref{modeCHARGE}) and (\ref{U1KM}),
\begin{equation}
[ {\cal H}^i_c \ , j^k_{-m} ] = \frac{\pi v_c}{\ell} m j^k_{-m}
\delta^{ik} \ .
 \label{HCj}
\end{equation}
Thus, comparing with (\ref{Ceigen}), the descendant levels in (the
diagonal basis of) the charge sector are obtained by adding energy to the
primary levels in steps of $\pi v_c/\ell$. The resulting finite-size
spectrum organizes into {\em conformal towers} of equally spaced energy
levels, each tower having a primary level $q^i$ as base:\footnote{The
invariance of ${\cal H}^i_c$ implies that its eigenstates fall into the
irreducible representations of affine $U(1)$. The notion of a {\em
conformal tower} is used interchangeably to denote these representations
{\em or} the corresponding set of spectral levels. The analogous remark
applies for ${\cal H}^i_s$, invariant under affine $SU(2)_1$.}
\begin{equation}
E^i_c(q^i, N^i_c) = \frac{\pi v_c}{\ell} \left\{ \frac{(q^i)^2}{4} -
\frac{1}{24} + N^i_c \right\} \ , \ \ \  N^i_c \in \N, \ i=1,2.
\label{EC}
\end{equation}
with
\begin{equation}
q^{{}^1_2} = Q \frac{e^\theta}{2} \pm \Delta Q \frac{e^{-\theta}}{2}
 \ , \ \ \ Q, \, \Delta Q \in \Z.
\end{equation}
Similarly, (\ref{modeSPIN}) and (\ref{SU2KM}) imply
\begin{equation}
[ {\cal H}_s \ , \vJ^k_{-m} ] = \frac{\pi v_s}{\ell} m \vJ^k_{-m} \ .
\label{HSj}
\end{equation}
A comparison with (\ref{Seigen}) yields the finite-size spectrum in the spin
sector, with two conformal towers ($j^i = 0,\onehalf$) per channel:
\begin{equation}
E^i_s(j^i, N^i_s) =
\frac{\pi v_s}{\ell} \left\{ \frac{1}{3}j^i(j^i+1) - \frac{1}{24} +  N^i_s
\right\} \ , \ \ \  N^i_s \in \N, \ i=1,2.
\label{ES}
\end{equation}
Putting $Q=\Delta Q = j^1 = j^2 =0$ and summing over the
channels in (\ref{EC}) and (\ref{ES}), we obtain the ground-state energy
$E^0 = E_c^0 + E_s^0$ with
\begin{equation}
E_j^0 = -\frac{\pi v_j}{12 \ell} \, \ \ j=c,s \ .
\label{CSenergy}
\end{equation}
As is well-known, the finite-size correction to the ground-state energy
$E^0 = E_L^0 + E_R^0$ of a conformally invariant Hamiltonian
(defined on a ring of circumference $L_x$) scales as
$E_{\LR}^0 = - \pi v c /12 L_x$ \cite{BCNA}, $v$ being the
velocity of the massless excitations. In the present case, we use a chiral
formulation and must accordingly compare (\ref{CSenergy}) with $E_L^0$ only.
Putting $L_x = 2\ell$, we obtain $c = 2$ in the charge and spin sectors,
respectively, yielding a total $c=4$. This is what we expect, since we are
using a $c=4$ representation of our original $c=2$ theory.

Given the finite-size spectrum, we now identify the corresponding scaling
dimensions. Using (\ref{SPEC}), together with (\ref{EC}) and
(\ref{CSenergy}), we have
\begin{equation}
\Delta^i_c = \frac{1}{4}\left( e^{\theta}
\frac{Q}{2} - (-1)^{i} e^{-\theta}\frac{\Delta Q}{2} \right)^2
\  + N^i_c \, , \ i=1,2 \ , \ N^i_c \in \N . \label{Cdim}
\end{equation}
Similarly, (\ref{SPEC}), (\ref{ES}), and (\ref{CSenergy}) imply for the
scaling dimensions in the spin sector:
\begin{equation}
\Delta^i_s = \frac{1}{3}j^i(j^i+1) + N^i_s \, , \ j^i=0,\onehalf, \
\ i=1,2 \ , \ \ \ N^i_s \in \N .  \label{Sdim}
\end{equation}
The dimension of a composite operator is thus given by
\begin{equation}
\Delta = \Delta^1_c + \Delta^2_c + \Delta^1_s + \Delta^2_s.
\label{comp}
\end{equation}
We recognize the scaling dimensions in (\ref{Cdim}) as
those of a $U(1)$ Gaussian theory represented by free bosons with
periodicity $\phi = \phi + 2 \pi {\cal R}$, with ${\cal R}
= e^{-\theta}/\sqrt{2}$
\cite{KB}. The same structure is hidden in (\ref{Sdim}). Substituting
$Q^{\pm} = Q^1_{ \uparrow} - Q^1_{ \downarrow} \pm ( Q^2_{ \uparrow} -
Q^2_{ \downarrow})$ for $Q$ and $\Delta Q$ respectively, choosing
$e^{-\theta} = 1 $ and using (\ref{connection}), (\ref{Cdim})
gives (\ref{Sdim}). For this special value of the periodicity
(${\cal R} = 1/\sqrt{2}$), it is known that the symmetry is enhanced
to $SU(2)$ \cite{Ginsparg}, and the scaling dimensions become those of
an $SU(2)$ $k=1$ WZW model.

It may here be worthwhile to recall the key elements of the analysis:
Equation (\ref{comp}) gives the spectrum of {\em chiral
dimensions} of the analytically continued theory in the full complex
plane. As we pointed out above, {\em boundary scaling
dimensions} in the half-plane correspond to subsets of this
spectrum, with one subset for each particular boundary condition.
By the ``folding procedure'' in (\ref{folding}), the subset corresponding
to the boundary condition (\ref{BC}) exactly coincides with the set
of {\em bulk dimensions} of the original theory in the full plane.
To pick out these dimensions from (\ref{comp}) we must first identify the
rule for assigning values to the quantum numbers $(Q, \Delta Q,
j^1, j^2)$ in presence of the boundary condition (\ref{BC}). By the
equivalence with the bulk problem, this is the same as identifying the
selection rule governing the bulk energy spectrum of ${\cal H}^*_{TL}$.

Assuming that the electron-electron interaction does not obstruct the
choice of particle number $(Q_{L\uparrow}, Q_{L\downarrow},
Q_{R\uparrow}, Q_{R\downarrow})$, (\ref{Q}) and (\ref{connection})
imply the {\em Luttinger liquid selection rule:}
\begin{equation}
\left\{ \begin{array}{rcl}
j^1 & = & \frac{1}{4} (Q + \Delta Q) \ \mbox{mod} \ 1 \\
j^2 & = & \frac{1}{4} (Q - \Delta Q) \ \mbox{mod} \ 1.
\end{array} \right.
\label{selectionI}
\end{equation}
Remember that $Q \pm \Delta Q$ is even, which is
consistent with the allowed values of $j^i$.
At this point we would also like to point out that this selection rule
includes an implicit relation between the two charge channels $1$ and $2$:
The definition of $q^i$ in terms of $Q$ and $\Delta Q$ only permits certain
combinations of the $q^1$ and $q^2$ conformal towers. In contrast, the
conformal towers in the spin sector are not constrained by such a relation.

The scaling dimensions of the possible boundary operators are now
obtained from (\ref{comp}) by feeding into (\ref{Cdim}) and
(\ref{Sdim}) the allowed values of $(Q, \Delta Q, j^1, j^2)$
according to (\ref{selectionI}). Not all of these operators appear,
though, in the effective theory describing the scaling region.
In general, this can be written as an expansion
\begin{equation}
{\cal H} = {\cal H}^* + \sum_k g_k {\cal O}_k(0) \, ,
\label{expansion}
\end{equation}
where ${\cal H}^*$ is the fixed point Hamiltonian, and $g_k$ and
${\cal O}_k$ are the associated scaling fields and boundary operators.
Corrections to scaling are produced by the irrelevant
operators, and these must be invariant under the continuous symmetries
of ${\cal H}$ (as must any relevant operators
in absence of external perturbations). Applied to our case,
${\cal H}$ = ${\cal H}^*_{TL} + \sum_k g_k {\cal O}_k(0)$,
invariant under chiral $U(1) \times U(1) \times SU(2)_1 \times SU(2)_1$.
In other words, charge and spin are conserved separately in the two
channels, implying $Q=\Delta Q = 0$ and $j^1 = j^2 = 0$ for ${\cal O}_k$.
Hence, the only boundary operators appearing at the fixed point are
descendants of the identity.
This trivial content of correction-to-scaling operators could of
course have been predicted directly from symmetry arguments, without
invoking the finite-size spectrum. However, having the spectrum in
hand, {\em including} the selection rule in (\ref{selectionI}), we can
attack the more challenging problem of electron-impurity scattering.

Before doing so, however, we shall compare our spectrum and
selection rule to those from an exact {\em Bethe Ansatz} analysis
of the Hubbard chain (of which ${\cal
H}^{*}_{TL}$ is the fixed point theory). As was first shown by Woynarovich
\cite{Woynarovich} (see also \cite{FK,KY}), the {\em Bethe Ansatz}
spectrum
of a Hubbard model on a finite ring also
organizes into conformal towers. Away from half-filling:
\begin{equation}
E - E_0 \sim \frac{2\pi v_c}{N} (\Delta_c^+ + \Delta_c^- ) +
\frac{2\pi v_s}{N} (\Delta_s^+ + \Delta_s^- )  + O(\frac{1}{N^2}).
\label{Wspec}
\end{equation}
Here N is the number of sites on the ring, $v_c$ and $v_s$ are model
dependent charge and spin velocities, and $\Delta_c^{\pm}$ and
$\Delta_s^{\pm}$ are given by
\bml
\begin{eqnarray}
\Delta_c^{\pm} & = & \frac{1}{2}\left(\frac{I_c}{2\xi_c} \pm \xi_c(D_c
+ \frac{D_s}{2}) \right)^2 + N_c^{\pm} \\
\Delta_s^{\pm} & = & \frac{1}{4}\left(I_s - \frac{I_c}{2} \pm D_s
\right)^2 + N_s^{\pm} ,
\end{eqnarray}
\label{BAdim}
\eml
with $N^{\pm}_c,N^{\pm}_s \in \N$.
As in the conformal approach, the positive integers $N^{\pm}$ label
``particle-hole excitations'' (although the notion of a ``particle''
or ``hole'' in a {\em Bethe Ansatz} basis is different from that
used here).
The parameter $\xi_c$ is a non-universal function of the microscopic
parameters, while $(I_c, I_s, D_c, D_s)$ are quantum numbers subject
to the {\em Bethe Ansatz selection rule}
\begin{equation}
\label{selectionBA}
\left\{ \begin{array}{rcll}
D_c & = & \frac{1}{2} (I_c + I_s) & \ (\mbox{mod} \ 1) \\
D_s & = & \frac{1}{2} I_c & \ (\mbox{mod} \ 1),
\end{array} \right. \ \ \ \ \ \ 2D_c, 2D_s, I_c, I_s \in {\Z}.
\end{equation}

To match our result in (\ref{Cdim}) and (\ref{Sdim}) to that in (\ref{BAdim}),
we note that the dressed
charge $\xi_c$ and our parameter $e^{\theta}$ both
measure the strength of the Hubbard interaction. Explicitly, using
the parameterization in Sec.~\ref{section2}, we have that
$e^{\theta} = [1+U/(2\pi t \, \mbox{sin}(ak_F))]^{1/4}$, while
$\xi_c = \sqrt{2} [1-U/(8 \pi t \, \mbox{sin}(ak_0))]$ \cite{FK},
$k_0$ playing the role of a ``Fermi momentum'' in the {\em Bethe Ansatz}
formalism. Putting $k_0 = k_F$, it follows that
$ \xi_c = \sqrt{2} e^{-\theta} + O[(U/t)^2]$. To lowest order in $U/t$,
$\Delta^1_c + \Delta^2_c$ and $\Delta^1_s + \Delta^2_s$ in (\ref{Cdim})
and (\ref{Sdim}) indeed exhaust the same {\em combined} spectrum of scaling
dimensions as $\Delta_c^{+}+\Delta_c^{-}$ and $\Delta_s^{+}+\Delta_s^{-}$
in (\ref{BAdim}). This can be seen from the following analysis.

Let us first generate all combinations of scaling dimensions according to
the Luttinger Liquid selection rule (\ref{selectionI}) and show that they
equal allowed combinations of scaling dimensions according
to the {\em Bethe Ansatz} selection rule (\ref{selectionBA}). We may
concentrate on $N^i_c = N^i_s = 0$, as higher levels can be trivially
mapped once the relation for $N^i_c = N^i_s = 0$ has been established.
Moreover, if we formally set $Q=I_c$ and use $\xi_c = \sqrt{2}
e^{-\theta}$, we see that it is sufficient to check that
$\Delta Q$ and $4 D_c + 2 D_s$ span the same range of integers when
$\Delta^1_s + \Delta^2_s$ = $\Delta_s^{+}+\Delta_s^{-}$.
Whether this correspondence is possible or not
depends on the selection rule.

Consider $Q=2n$, $n \in \Z$. By construction, $\Delta Q$ is even, and
from (\ref{Sdim}) and (\ref{selectionI}) it follows that for
$\Delta_s^1 + \Delta_s^2 = 0$ we have
$\Delta Q = 4m \ \backslash 4m+2\backslash$,
$m \in \Z$, for $n$ even $\backslash$odd$\backslash$.
This is exactly reproduced in the {\em Bethe Ansatz}
spectrum by choosing $D_s=0$ (which is allowed as $I_c$ is even)
and $I_s=n$. According to the selection rule, the allowed values
of $D_c$ are $p \ \backslash p + \onehalf\backslash$,
$p \in \Z$, for $n$ even $\backslash$odd$\backslash$, i.e.
$4 D_c + 2 D_s$ gives the same range of integers as $\Delta Q$.
The other possibility, $\Delta_s^1 + \Delta_s^2 = \onehalf$, is
analogously combined with the above values of $\Delta Q$ shifted by 2,
which is reproduced by choosing $D_s=0$, $I_s=n+1$, and
$D_c = p + \onehalf \ \backslash p \backslash$, $p \in \Z$.

Next, consider $Q = 2n+1$, $n \in \Z$. Then $\Delta_s^1 + \Delta_s^2 =
\frac{1}{4}$ and $\Delta Q$ spans all odd integers. Choosing $I_s = n$ and
$D_s = \pm\onehalf$ is consistent with (\ref{selectionBA}) and gives
$\Delta_s^+ + \Delta_s^- = \frac{1}{4}$. For $I_s$ even
$\backslash$odd$\backslash$, the allowed values of $D_c$ are $p + \onehalf
\ \backslash p \backslash$, $p \in \Z$. In both cases, $4 D_c + 2 D_s$
spans all odd integers as required. This completes the first part of our
comparison.

In order to prove that the Luttinger liquid and {\em Bethe Ansatz} spectra
are identical, we must also check that any other allowed combination of quantum
numbers of the latter only reproduce energy levels already obtained.
(Note that we only compare energy levels and not their degeneracies.)
So far we have exhausted all combinations of $I_c$ and $D_c$. By also
allowing $I_s$
and $D_s$ to take any permissible value, one can show that all energy levels
in the {\em Bethe Ansatz} solution for $N_c^\pm = N_s^\pm = 0$ correspond
to allowed combinations of
$\Delta_c^1 + \Delta_c^2$ and $\Delta_c^1 + \Delta_c^2$ with
$N_c^i=0$ and $N_s^i \in \N$, respectively. The calculation is
straightforward, so we leave it out for brevity. As before, it is trivial
to extend the mapping to include $N_c^\pm, N_s^\pm \in \Zplus$, and thus
the desired result follows.

\subsection{Coset construction: Boundary operators in presence of ${\cal H}_F$}

Having established the spectrum of ${\cal H}_{TL}^*$, we now include
the electron-impurity term
\begin{equation}
{\cal H}_F = \lambda [ \vJ_L^1(0) + \vJ_L^2(0)]\cdot \vS
\end{equation}
and explore how this interaction affects the boundary scaling
dimensions. For this purpose, it is useful to rewrite the
Hamiltonian ${\cal H}^*_{TL} + {\cal H}_F$ in terms of the total
spin current $\vJ = \vJ^1_L + \vJ^2_L$:
The chiral $SU(2)_1 \times SU(2)_1 $ symmetry of ${\cal H}^*_{TL}$
gets broken by ${\cal H}_F$, whereas the diagonal subgroup $SU(2)_2$,
generated by $\vJ$, remains as a symmetry (cf. our discussion in
Sec.~\ref{section2}).

In the conformal field theory formalism it is sufficient to work at
the level of representations of the $SU(2)_k$ algebras. Given a
direct product of two irreducible representations (conformal towers)
of $SU(2)_1$, we thus ask how the states reappear using
representations of $SU(2)_2$. The answer is immediately obtained
from the GKO coset construction \cite{GKO}: Products of two
$SU(2)_1$ conformal towers $(j^1)_1 \times (j^2)_1$
decompose into products of $SU(2)_2$ and Ising model conformal
towers $(j)_2 \times (\phi)$ according to
\begin{eqnarray}
(0)_1 \times (0)_1 &=& (0)_2 \times ({\1}) +
(1)_2 \times (\epsilon) \nonumber \\
(0)_1 \times (\onehalf)_1 &=& (\onehalf)_2 \times (\sigma) \nonumber \\
(\onehalf)_1 \times (\onehalf)_1 &=& (0)_2 \times (\epsilon)
+ (1)_2 \times ({\1}) \, . \label{GKO}
\end{eqnarray}
(Note that $(\onehalf)_1 \times (0)_1$ is degenerate with $(0)_1 \times
(\onehalf)_1$.)
Here $j^i = 0,\onehalf$ and $j=0,\onehalf,1$ label the conformal
towers of $SU(2)_1$ and $SU(2)_2$, respectively, while $\phi = {\1}$
(identity), $\sigma$ (order parameter), and $\epsilon$ (energy density)
label the three conformal towers of the Ising model ($c=\onehalf$).
The scaling dimensions in the $SU(2)_2$ sector are given by
\begin{equation}
\Delta_S = \frac{1}{4}j(j+1) + N_S, \ \ N_S \in \N \label{SU2dim}
\end{equation}
while
\begin{equation}
\Delta_{Ising} = \left\{
\begin{array}{cccc}
0 & + & N & ({\1}) \\
\frac{1}{16} & +& N & (\sigma)  \\
\frac{1}{2} & + & N & (\epsilon)
\end{array}  \right . \ \ \ N \in \N \label{Isingdim}
\end{equation}
are those of the Ising sector.

All primary states in the new representation are now labeled
by $(Q, \Delta Q, j, \phi)$, with the new selection rule obtained by
combining (\ref{selectionI}) and (\ref{GKO}):
\begin{equation}
(j,\phi) = \left\{ \begin{array}{cccl}
 (0,\1) & \mbox{or} & (1,\epsilon) & \ \ \ \ \ Q \ \mbox{even}, \ \onehalf
 (Q+\Delta Q) \ \mbox{even} \\
 (0,\epsilon) & \mbox{or} & (1,\1) & \ \ \ \ \ Q \ \mbox{even}, \ \onehalf
 (Q+\Delta Q) \ \mbox{odd} \\
 (\onehalf, \sigma) & &            & \ \ \ \ \ Q \ \mbox{odd}.
\end{array} \right.
\label{selectionII}
\end{equation}
At the Hamiltonian level, the conformal embedding $SU(2)_1 \times
SU(2)_1 \rightarrow SU(2)_2 \times Ising$ implies that the spin part
of ${\cal H}_{TL}$
decomposes into a sum of an $SU(2)_2$
Sugawara and a free Majorana Hamiltonian:
\begin{equation}
\tilde{\cal H}_s = \frac{v_s}{2 \pi}
\int_{-\ell}^{\ell} dx \, \biggl\{ \frac{1}{4} \no{\vJ(x) \cdot \vJ(x)} +
\eta_L(x) i \frac{d}{dx} \eta_L(x) \biggr\} \, , \label{confemb}
\end{equation}
with $\eta_L$ a left-moving Majorana fermion.
This
enables us to absorb the electron-impurity term ${\cal H}_F = \lambda
\vJ(0) \cdot \vS$ into $\tilde{\cal H}_s$ by making the canonical
transformation
\begin{equation}
\vJ(x) \rightarrow \vJ(x) + \vS \delta(x) \ ,
\label{Jtransf}
\end{equation}
and choosing $\lambda = \lambda^* \equiv \frac{v_s}{4\pi}$.
In this scheme $\lambda^*$ defines the (non-universal) fixed point coupling:
the elimination of ${\cal H}_F$ by (\ref{Jtransf}) restores
translational (and thereby conformal) invariance, implying a fixed
point theory.

The redefinition of the total current in (\ref{Jtransf}) changes the
rule for coupling conformal towers. Effectively, (\ref{Jtransf})
adds an extra spin-$\onehalf$ degree of freedom to the $SU(2)_2$
sector, leaving the $U(1)$ and Ising sectors intact. As a result,
the $SU(2)_2$ conformal towers $(j)_2$ get replaced,
according to
\begin{equation}
(0)_2 \rightarrow (\onehalf)_2 , \ \ \ (\onehalf)_2 \rightarrow
(0)_2 \ \mbox{or} \ (1)_2 , \ \ \ (1)_2 \rightarrow (\onehalf)_2 .
\label{gluing}
\end{equation}
These are the $SU(2)_2$ {\em fusion rules}, see (\ref{SU2fusion}), describing
the effect of combining a tower of spin $j$ with a spin-$\onehalf$ tower.
Together with (\ref{selectionII}) this leads to the new selection rule
\begin{equation}
(j,\phi) = \left\{ \begin{array}{cccl}
 (\onehalf,\1) & \mbox{or} & (\onehalf,\epsilon) & \ \ \ \ \ Q \ \mbox{even} \\
 (0,\sigma) & \mbox{or} & (1,\sigma) & \ \ \ \ \ Q \ \mbox{odd},
\end{array} \right.
\label{selectionIII}
\end{equation}
which governs the finite-size spectrum of ${\cal H}_{TL}^* + {\cal H}_F$,
or equivalently, the spectrum of ${\cal H}_{TL}^*$ with a modified boundary
condition at $x=0$ representing the impurity spin.

However, to obtain the scaling dimensions related to a
boundary condition, we recall from the beginning of Section~\ref{section3}
that these are in one-to-one correspondence to the finite-size spectrum of
the Hamiltonian with that boundary condition applied at {\em both} ends.
In terms of the Hamiltonian (\ref{newSTAR}), defined on $x \in [0,\ell]$,
this corresponds to having impurity spins at both ends of the space interval.
Passing over to the chiral formulation (\ref{Schiral}), we thus see that the
boundary scaling dimensions of ${\cal H} = {\cal H}^*_{TL} + {\cal
H}_F$ are in one-to-one correspondence to the energy levels of the
{\em auxiliary Hamiltonian}
\begin{equation}
{\cal H}' = {\cal H}^*_{TL} + \lambda \int_{-\ell}^{\ell} dx \, \vJ(x)
\cdot [ \vS_1 \delta (x) + \vS_2 \delta (x- \ell)]. \label{aux}
\end{equation}
The two impurity terms are removed at the fixed point by the canonical
transformation
\begin{equation}
\vJ(x) \rightarrow \vJ(x) + \vS_1 \delta(x) + \vS_2
\delta(x-\ell) , \label{Jtransf2}
\end{equation}
by which {\em two} extra spin-$\onehalf$ degrees of freedom are added to the
$SU(2)_2$ sector. The fusion hypothesis \cite{AL1} suggests that the effect
is described by {\em two} repeated fusions of the spin $j$ conformal towers
with a $j=\onehalf$ tower. The new selection rule that emerges is
readily extracted, using (\ref{gluing}) twice to replace the $SU(2)_2$ towers
in (\ref{selectionII}) by the resulting set of conformal towers. One finds,
\begin{equation}
(j,\phi) = \left\{ \begin{array}{ll}
 (0 \ \mbox{or} \ 1,\1 \ \mbox{or} \ \epsilon)  & \ \ \ \ \ Q \ \mbox{even} \\
 (\onehalf,\sigma)                              & \ \ \ \ \ Q \ \mbox{odd}.
\end{array} \right.
\label{selectionIV}
\end{equation}
As before, $\Delta Q$ has the same parity as $Q$, i.e. they are both even
or both odd.

Summarizing, the spectrum of boundary dimensions in the presence of
${\cal H}_F$ is obtained from
\begin{equation}
\Delta = \Delta^1_c + \Delta^2_c + \Delta_S + \Delta_{Ising},
\label{newcomp}
\end{equation}
using (\ref{selectionIV}) to insert the allowed combinations of
$(Q, \Delta Q, j, \phi)$ into the expressions for $\Delta^i_c \
(i=1,2),
\Delta_S$ and $\Delta_{Ising}$ in (\ref{Cdim}), (\ref{SU2dim}),
and (\ref{Isingdim}), respectively.

\section{Critical Behavior}
\label{section4}

We now turn to the question of the {\em impurity critical behavior}, exploiting
the result derived in the previous section. To do so requires two steps.
First, the {\em symmetry preserving} boundary operators must be identified
from the spectrum in (\ref{newcomp}). With these in hand,
one then selects the operator of lowest dimension to
perturbatively calculate the leading finite-size corrections to the fixed
point theory. Treating the (inverse) length as a temperature
variable allows for the calculation of the finite-$T$ scaling of
physical response functions, such as the
impurity contribution to the specific heat and magnetic susceptibility.

It is tempting to make a shortcut and refer to the logic of
our procedure to
conclude that the impurity critical behavior must be that
of the two-channel Kondo problem: The selection rule in (\ref{selectionIV}),
describing the renormalized boundary condition due to the impurity,
descends from the Luttinger liquid selection rule in (\ref{selectionI}) via
the coset construction and ``double fusion''. This selection rule
in turn is nothing but the free-electron selection rule in disguise,
``rotated'' to a diagonal basis in the charge sector.
One thus infers that the interacting problem cannot
differ in essence, provided that the charge sector does not
contribute a (non-trivial) leading correction-to-scaling boundary
operator (which, by chiral charge conservation, is excluded).
Although this line of reasoning is essentially correct, it is
instructive to carry out the analysis ``by hand'', and
observe how two-channel Kondo physics emerges in the scheme
proposed here. In fact, certain novel features appear: the
leading correction-to-scaling boundary operator {\em (LCBO)} is
found to be {\em unique} at the fixed point, suggesting a universal
Wilson ratio in the low-temperature,
strong coupling limit. Also, by working out the solution
explicitly in the new scheme we gain a number of important
insights, to be exploited when extending the analysis to Kondo
scattering in a Luttinger liquid (Sec.~\ref{section5}).

\subsection{Finite-temperature scaling}

To provide some background, let us briefly review the finite-temperature
scaling approach to the problem, including the rather exotic
mechanism that brings the {\em irrelevant} boundary operators to
center stage. For more details, we refer to Refs. \cite{AL1,Lrev}.

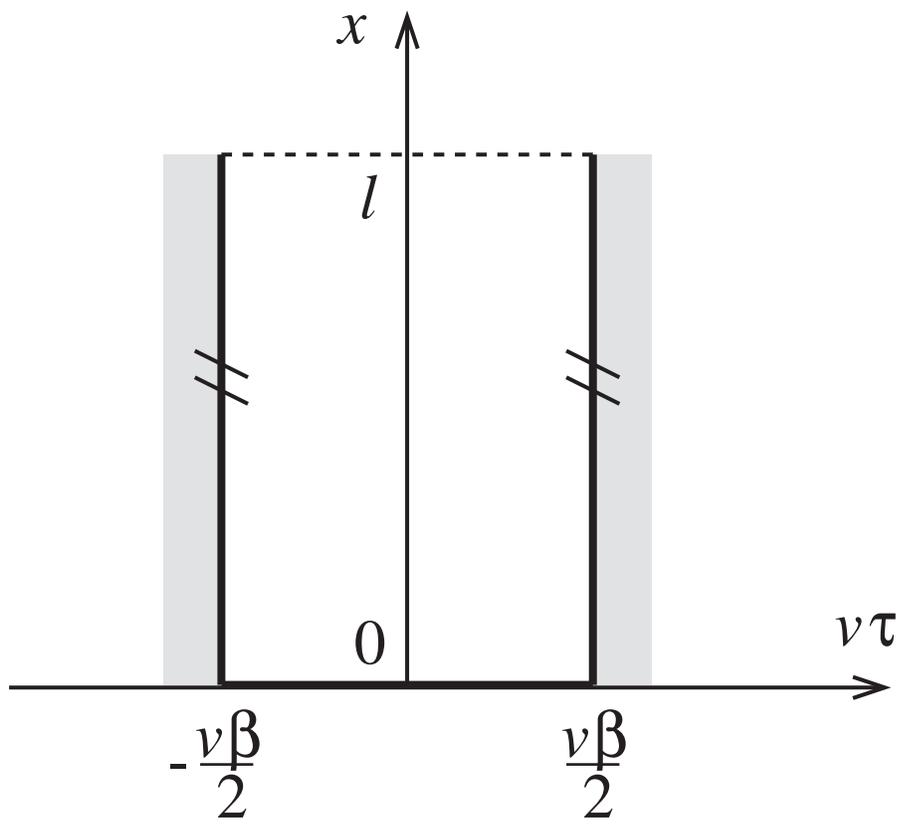
\begin{figure}
\caption{Finite-temperature geometry with periodic boundary condition in
imaginary time $\tau$. The temperature $T$ is identified as $1/\beta$ and
the circumference $v\beta$ of the cylinder satisfies $v \beta \ll \ell$.
For $\ell \rightarrow \infty$, this geometry is related to the
zero-temperature geometry in Fig.~4b via the conformal map
$w = (v\beta/\pi)\mbox{arctan}(z)$.}
\label{fig:finiteT}
\end{figure}

Consider the (folded) system confined to a spatial interval $[0, \ell]$ and
in the presence of an external magnetic field $ h$. By treating
temperature $T$ as an inverse length $1/\beta$, we may extract the
low-temperature thermodynamics via finite-size scaling on a cylinder with
circumference $v\beta$. A convenient choice is to map the upper
half-plane $\Cplus = \{\im z > 0 \}$ ($T=0$ geometry in
Fig.~\ref{fig:strip-halfplane}b) onto a
cylinder $\Gammaplus = \{ w = ({v\beta} / {\pi}) \,
\mbox{arctan} z \}$ (finite-$T$ geometry in Fig.~\ref{fig:finiteT}).
At finite temperature the free energy separates into two pieces, describing
the bulk and the impurity, respectively:
\begin{equation}
F(\beta,  h, \lambda_I) = 2 \ell f_{bulk} (\beta,  h) +
f_{imp}(\beta,  h, \lambda_I) \, , \label{Free}
\end{equation}
with $\lambda_I$ the leading irrelevant boundary scaling
field.\footnote{Note that we omit the irrelevant bulk fields in
(\ref{Free}). This can safely be done: {\em If} there is a
correlation between irrelevant bulk and
boundary operators, the corresponding term in a perturbative expansion of
$f_{imp}$ in $T$ can be shown to produce a subleading contribution compared to
that coming from the leading irrelevant boundary operator alone. On the
other hand, a bulk irrelevant field may produce nonanalytic
corrections to $f_{bulk}$, but this is of no concern to us here.}
By writing $2\ell$ instead of $\ell$ in (\ref{Free}), we let $f_{bulk}$
refer to the free-energy density of the original (unfolded) system;
cf. Figs.~\ref{fig:fold}a and~\ref{fig:fold}b and the remark after
(\ref{newFORWARD}).

{}From the standard finite-size scaling hypothesis \cite{Privman}, it
follows that the reduced free-energy density of a critical theory, defined
on a cylinder with circumference $v \beta$, scales as
\begin{equation}
\beta F / \ell = f_0 v\beta + Y(h\beta) /v \beta + \ldots \, ,
\end{equation}
where the universal amplitude $Y_{bulk}(0) = -\pi c/6$ \cite{BCNA}.
In our case, the critical theory is a sum of two conformal theories,
each with $c=2$, and with effective velocities $v = $ $v_{c}$ and $v_{s}$,
respectively. Moreover, as the magnetic field $h$ only couples to the
spin sector, we may simply add the contributions from the two sectors:
\begin{equation}
f_{bulk}(\beta,h) = E_{bulk} - \frac{\pi}{6v_c\beta^2}
  + \frac{1}{2 v_s\beta^2} Y_s(h\beta) + \ldots \, , \label{bulkscale}
\end{equation}
with $E_{bulk}$ a non-universal quantity and $Y_s(0) = -\pi/3$.
Putting $\beta = 1/T$ one thus obtains for the bulk specific heat:
\begin{equation}
C_{bulk}(T) = -T\frac{\partial^2 f_{bulk}}{\partial T^2} =
\frac{\pi}{3}(\frac{1}{v_c}+\frac{1}{v_s})T + \ldots \, , \label{Cbulk}
\end{equation}
with ``$\ldots$'' denoting subleading terms as $T \rightarrow 0$.
The bulk magnetic susceptibility can similarly be obtained from $f_{bulk}$
by expanding $Y_s$ to second order in $h/T$,
and one finds
\begin{equation}
\chi_{bulk}(T) = -\frac{\partial^2 f_{bulk}}{\partial
h^2} \Big|_{h=0} = \frac{1}{2 \pi v_s} + \ldots \label{chibulk}
\end{equation}

In the same manner as for $f_{bulk}$ one can write down a finite-size scaling
ansatz for the impurity part of the free energy:
\begin{equation}
f_{imp}(\beta,  h, \lambda_I) = E_{imp} + \frac{1}{\beta}Y_{imp}(
h\beta, \lambda_I\beta^y) + \dots \, , \label{impscale}
\end{equation}
with $E_{imp}$ non-universal. The exponent $y$ is an RG eigenvalue,
connected to the dimension $\Delta$ of the leading irrelevant
boundary operator by $y=1-\Delta$. As the boundary scaling field $\lambda_I$
may couple to both the charge and the spin sector ($\Delta = \Delta_c +
\Delta_s$), $Y_{imp}$ is not just a sum of a charge and a spin part.
This will not be important, though, as we will later calculate $f_{imp}$
without any assumptions about $Y_{imp}$. However, for $\lambda_I = 0$ there
is only one scaling field $h$ and it couples only to the spin sector.
The $h$-dependence of $Y_{imp}$ is therefore given by the same universal
function as in a theory with a single velocity.

{}From the scaling form of $Y_{imp}$ one expects non-analytic temperature
terms in $f_{imp}$ for non-integer $y$ and $\lambda_I \neq 0$.
Remarkably, the impurity specific heat and susceptibility vanish
identically when $\lambda_I = 0$, so these non-analytic terms become
dominant.
The mechanism behind this unusual behavior is particularly
transparent within the conformal field theory approach: Since, {\em
at the fixed point}, the
impurity spin has been absorbed in the spin current $\vJ$, the fixed point
theory has lost all memory of the impurity.
Specifically, the total magnetization
\begin{equation}
\int_{-\infty}^{\infty} dx \cf{J^z (x)} = \frac{\partial
F(\beta, h,\lambda_I = 0)}{\partial h}
\end{equation}
is insensitive to the impurity, and thus, from (\ref{impscale}):
\begin{equation}
\frac{\partial Y_{imp}(h\beta, \lambda_I = 0)}{\partial(h\beta)} = 0
\, .
\end{equation}
This implies that exactly {\em at} the fixed point
\begin{equation}
\chi_{imp}(\lambda_I=0) =
- \frac{\partial^2 f_{imp}}{\partial h^2}\Big|_{h=0} = -\beta
\frac{\partial^2 Y_{imp}(h\beta, \lambda_I = 0)}
{\partial(h\beta)^2}\Big|_{h=0} = 0 \, ,
\end{equation}
and analogously
\begin{equation}
\label{Cimp_crit}
C_{imp}(\lambda_I=0) = -T \frac{\partial^2 f_{imp}}{\partial T^2} = 0.
\end{equation}
To ``put back'' the effect of the impurity, the leading
irrelevant boundary operator must be added, and this operator then
produces the {\em dominant} term in the scaling in temperature.

\subsection{Boundary operators}
\label{section4B}

Following the procedure outlined in the beginning of this section,
we first pick out the scaling dimensions in (\ref{newcomp}) corresponding
to the symmetry preserving operators.

It is easiest to start in the charge sector. Chiral $U(1)$ invariance
implies charge conservation in each channel, hence $Q = \Delta Q = 0$.
The selection rule in (\ref{selectionIV}) then implies that possible
boundary operators can appear only in the products of conformal towers
\begin{equation}
(Q, \Delta Q , j, \phi) = (0,0,0,\1), \ (0,0,1,\1), \ (0,0,0,\epsilon),
\
\mbox{and} \ (0,0,1,\epsilon). \label{towerproducts}
\end{equation}

The first and fourth structures in (\ref{towerproducts}) contain only
operators with integer scaling dimensions. As we shall see in the
next section, these produce Fermi-liquid like, analytic scaling in
temperature, and will be disregarded at this point. Turning to
the second and third structures, we note that
both contribute a {\em relevant} boundary operator, with
$\Delta < 1$.\footnote{An operator of scaling dimension $\Delta$ is relevant
if $D-\Delta > 0$. At a boundary, $D=1$, hence $\Delta < 1$
implies relevance.} The second possibility, $(0,0,1,\1)$, contains the
spin-1 primary field, call it $\vphi$, with $\Delta = \onehalf$.
However, the {\em total} spin is conserved, requiring all operators in
the $SU(2)_2$ sector to transform as singlets. This expels $\vphi$,
leaving only descendant operators in this sector with
$\Delta \geqslant \frac{3}{2}$. The third structure, $(0,0,0,\epsilon)$,
contains the Ising energy density as a primary field, also with
dimension $\Delta = \onehalf$. Is it also suppressed by symmetry, or do we
have to ``fine tune'' the theory to stay at the fixed point?
Consider the latter alternative. The operator $\epsilon$ originates from the
conformal embedding of $SU(2)_1 \times SU(2)_1$ into
$SU(2)_2 \times {\Z}_2$. The only parameter
in ${\cal H} = {\cal H}^*_{TL} + {\cal H}_F$ multiplying the $SU(2)_1$
currents is the spin velocity $v_s = v_F - g$. A ``fine-tuning''
scenario thus implies that we can stay at the fixed point only for
some privileged value of $g$, which, considering the known
solution for non-interacting electrons, must be $g=0$. Although not
excluded {\em a priori}, this is not a likely situation. Let us
instead explore whether there is any symmetry that suppresses
$\epsilon$.

The important symmetry to consider is invariance under channel
exchange: ${\cal E}: 1 \leftrightarrow 2$, under which $\vJ^1_L(x)
\leftrightarrow \vJ^2_L(x)$.
To test for ${\cal E}$, we connect
$\epsilon$ to the spin currents via the following observation:
The difference of $SU(2)_1$
currents, $\vJ^1_L(x) - \vJ^2_L(x)$, transforms in the adjoint (spin-1)
representation of {\em global} $SU(2)$ (with the
Schwinger term subtracted):
\begin{equation}
[J^{1a}_L(x) + J^{2a}_L(x) \ , \ J^{1b}_L(y) - J^{2b}_L(y) ] = i\epsilon^{abc}
(J^{1c}_L(x) - J^{2c}_L(x)) \delta (x-y).
\end{equation}
Knowing that the currents carry dimension 1, one is lead to the
identification
\begin{equation}
\vJ^1_L(x) - \vJ^2_L(x) \sim \vphi(x) \times \epsilon(x). \label{diff}
\end{equation}
This gives the correct assignment of spin (=1) and scaling dimension
$(\Delta_{\vphi} + \Delta_{\epsilon} = \onehalf + \onehalf = 1)$.
Now consider a channel exchange ${\cal E}$. According to (\ref{diff}),
\begin{equation}
{\cal E}: \vphi  \times \epsilon  \rightarrow - \vphi  \times
\epsilon . \label{exchange}
\end{equation}
A consistent conformal representation hence requires that we assign
evenness (oddness) to $\epsilon$ ($\vphi$) under ${\cal E}$, {\em or} vice
versa. Consider the first possibility: $\epsilon$ even and
$\vphi$ odd. Since ${\cal H}$ is invariant under ${\cal E}$,
$\vphi$ and all its descendants become suppressed.
The operator $\epsilon$ on the other hand is allowed, and the
required fine-tuning leaves us with a critical theory only at $g=0$,
according to the argument above.
The first descendant of $\epsilon$, $L_{-1}\epsilon$ with scaling
dimension $\frac{3}{2}$, here becomes the leading irrelevant operator.
But this is exactly the boundary operator that drives two-impurity
Kondo behavior \cite{AL2}, in contradiction to the {\em known}
two-channel behavior at $g=0$. Agreement with established results
in the non-interacting limit forces us to make the alternative
assignment under ${\cal E}$: $\epsilon$ odd and $\vphi$ even.
Now the $\epsilon$ conformal tower gets suppressed, with no fine
tuning necessary. Our analysis shows that there is no relevant boundary
operator present: The only candidates appearing in (\ref{towerproducts}),
$\vphi$ ($\Delta = \onehalf$) and $\epsilon$ ($\Delta
= \onehalf$) are expelled by symmetry. This yields the important
conclusion that the theory flows onto a {\em stable} fixed point in
the absence of external perturbations.

The leading irrelevant boundary operator, call it
${\cal O}_I$, is obtained by contracting $\vphi$ with the first $SU(2)_2$
raising operator $\vJ_{-1}$: ${\cal O}_I = \vJ_{-1} \cdot \vphi$,
with dimension $\frac{3}{2}$.
As expected, ${\cal O}_I$ is the same boundary operator that appears in the
analysis of the two-channel Kondo effect, using the ``charge-spin-flavor''
scheme \cite{AL1}, reviewed in Sec.~\ref{section2}. In that formulation,
however, the flavor sector contributes a second operator with dimension
$\frac{3}{2}$, call it ${\cal O}_{flavor}$. This is different from our
approach where the leading correction-to-scaling operator ${\cal O}_I$
is unique. Consistency between the two schemes requires that the scaling
field conjugate to ${\cal O}_{flavor}$ vanishes at the fixed point. Based
on the related case of ``underscreening'', it has been argued that this
indeed happens: When the impurity carries unit spin both the charge
and flavor sectors contribute correction-to-scaling operators with
the same dimension (=2) as the spin sector. Given a proper
regularization of the theory, one can show that the conjugate
charge and flavor scaling fields $\rightarrow 0$ at the trivial
$T>T_K$ fixed point. There is evidence that an analogous mechanism
is at play at the ``overscreened'' $T<T_K$ fixed point (with $s=\onehalf$),
leaving only the operator ${\cal O}_I$ effective. Our result
supports this picture, and allows for the definition of a universal
Wilson ratio in the limit of vanishing electron-electron interaction.

\subsection{Low-temperature thermodynamics}

The low-temperature thermodynamics driven by ${\cal O}_I$ has been
studied in \cite{AL1}. In the next section we shall review this analysis,
and extend it to theories with more than one velocity and boundary operators
of arbitrary dimensions. Here we only quote the result for the impurity
specific heat $C_{imp}$ and magnetic susceptibility $\chi_{imp}$:
\begin{eqnarray}
C_{imp}(T,\lambda_I) & = & \frac{\lambda_I^2 9 \pi^2}{v_s^3}
 T \ln (\frac{1}{\tau_0T}) + O(\tau_0T), \label{Cimp} \\
\chi_{imp}(T,\lambda_I) & = & \frac{\lambda_I^2 18}{v_s^3}
\ln (\frac{1}{\tau_0T}) + O[(\tau_0T)^0], \label{CHIimp}
\end{eqnarray}
with $\tau_0$ a short-time cutoff, playing the role of an inverse
``Kondo temperature''.
The scaling field $\lambda_I$ (conjugate to ${\cal O}_I$) appears with
the same power in
(\ref{Cimp}) and (\ref{CHIimp}). Thus, given the known bulk response
in (\ref{Cbulk}) and (\ref{chibulk}), one may form the $\lambda_I$-
and $T$-independent Wilson ratio
\begin{equation}
R_W \equiv
\frac{\chi_{imp} / \chi_{bulk}}{C_{imp} / C_{bulk}} =
\frac{4}{3} \left( 1 + \frac{v_s}{v_c} \right) =
\frac{8}{3}\left( 1-\frac{g}{v_F} + O[(\frac{g}{v_F})^2]\right).
\end{equation}
In the $g \rightarrow 0 \ (v_c, v_s \rightarrow v_F)$ limit we
recover the universal number $\frac{8}{3}$ obtained by Affleck
and Ludwig \cite{AL1} for the two-channel Kondo effect.

In the alternative charge-spin-flavor scheme for the two-channel Kondo effect,
the flavor sector also contributes a dimension-$\frac{3}{2}$ boundary operator.
This produces a second term in $C_{imp}$, proportional to the square
of a flavor scaling field $\lambda_{flavor}$. To obtain the Wilson ratio
$R_W = \frac{8}{3}$, one has to resort to the argument sketched
above, suggesting that $\lambda_{flavor} \rightarrow 0$ as one approaches
the fixed point. In contrast, the reduction to {\em one} scaling
field $\lambda_I$ is automatic in our approach.

The results in (\ref{Cimp}) and (\ref{CHIimp}) also apply to the
associated spin chain problem, with two neighboring sites coupled
antiferromagnetically with equal strength $J_0$ to a spin-$\onehalf$
impurity (cf. Fig.~\ref{fig:Heis} and the discussion in Sec.~\ref{section2}).
This corroborates the result in \cite{CGS}, and shows that the closing
of an open chain with its two ends coupled symmetrically to the same
impurity does not affect the scaling behavior: both the open and closed
chain exhibit the same ``two-channel Kondo physics'' \cite{EA}.
This, from a naive point of view, is somewhat surprising:
The two spins at the endpoints of the open chain are locked into an $S=1$
state via the interaction with the impurity. This causes overscreening of
the impurity spin, resulting in the two-channel Kondo behavior. Upon
closing the chain the endpoint spins get mutually coupled via an
antiferromagnetic exchange, call it $J$. As $J$ is increased above $J_0$
one may have guessed that the local $S=1$ state gets destabilized, thus
inhibiting overscreening, and leading to a decoupling of the impurity
spin (``Curie behavior''). This does not happen however. The results in
(\ref{Cimp}) and (\ref{CHIimp}) are valid independent of the relative
strengths of the bare couplings $J_0$ and $J$: The low-temperature fixed
point is {\em stable}, attracting a flow for arbitrary initial (bare)
values of the couplings $J$ and $J_0$. What {\em does} change, however,
is the temperature scale at which one reaches this fixed point. For example,
if the ratio $J_0/J$ is very small, the Kondo temperature will also be small,
and one has to go to very low temperatures to see the ``two-channel behavior''.

\subsection{Symmetry breaking perturbations}

It is instructive to study the stability of the forward scattering fixed
point against symmetry breaking perturbations. Consider first an asymmetric
electron-impurity interaction,
\begin{equation}
{\cal H}_F' = \lambda_L \vJ_L(0) \cdot \vS + \lambda_R \vJ_R(0) \cdot
\vS \, , \label{asymmetric}
\end{equation}
with $\lambda_L$ and $\lambda_R$ arbitrary. This may be expressed as
${\cal H}_F' = {\cal H}_F + {\cal H}_{pert}$ where ${\cal H}_F$ is
the symmetric electron-impurity interaction (with $\lambda = \onehalf
(\lambda_L + \lambda_R)$) studied above, and
\begin{equation}
{\cal H}_{pert} = \frac{1}{2}(\lambda_L - \lambda_R) \left[ \vJ^1_L(0) -
\vJ^2_L(0) \right] \cdot \vS . \label{perturbation}
\end{equation}
${\cal H}_{pert}$ breaks invariance under channel exchange, and
using (\ref{diff}) one has
\begin{equation}
{\cal H}_{pert} \sim  \left[ \vphi(0) \times
\epsilon(0) \right] \cdot \vS. \label{diff2}
\end{equation}
It follows that $\epsilon$ enters as an allowed boundary operator,
being contained in the spectrum (\ref{towerproducts}). Having
dimension $\Delta = \onehalf$, it is relevant, and destabilizes the
symmetric forward scattering fixed point.\footnote{Note that the spin-1
field $\vphi$ is odd under time reversal $({\cal T})$, and hence
cannot appear alone, but only as a piece of a composite operator
that respects ${\cal T}$. As can be seen from (\ref{diff}),  $\epsilon$ is
even under ${\cal T}$, and is thus allowed. Also note that exactly {\em at} the
fixed point, the scaling field conjugate to $\epsilon$ vanishes, and
the invariance under $\epsilon \rightarrow -\epsilon$ of the
critical Ising model is recovered {\em (Kramers-Wannier duality)}.}
This is similar to adding a
flavor anisotropic perturbation to the two-channel Kondo
interaction, using a charge-spin-flavor scheme \cite{ALPC}. In this
case the flavor sector contributes a dimension-$\onehalf$ operator
which takes the system to a one-channel (Fermi liquid like) fixed point.
Again, equivalence between the two schemes requires $\epsilon$ to
perform the same way. An interesting question is how the scaling
region about this one-channel fixed point is influenced by the
electron-electron interaction. However, we shall not pursue this problem here.

Instead, let us study the effect of adding electron backscattering
on the impurity, that is, adding ${\cal H}_B$ in (\ref{B}) to
${\cal H}_F$. ${\cal H}_B$ breaks both chiral SU(2) {\em and} U(1)
invariance. In particular, this means that the numbers of left and right
moving particles are not conserved separately. As a consequence,
$\Delta Q \equiv Q_L - Q_R$ is no longer restricted to zero, and the
charge sector can now make non-trivial contributions to the spectrum
of boundary operators. The lowest-dimension operator with $\Delta Q
\neq 0$ allowed by the selection rule in (\ref{selectionIV})
is obtained from
\begin{equation}
(Q, \Delta Q, j, \phi) = (0, \pm 2, 0 , \1), \label{Brel}
\end{equation}
and has dimension $\Delta = \onehalf e^{-2\theta} \leqslant \onehalf $.
Any $|\Delta Q | > 2$ will produce an irrelevant operator, and hence
the composite primary operator from (\ref{Brel}) is the
unique relevant operator generated by breaking chiral U(1)
invariance. (Note that the breaking of chiral SU(2) invariance is
already effected by ${\cal H}_F$.) As expected, ${\cal H}_B$ is thus
a {\em relevant perturbation} that pulls the system away from the
forward scattering fixed point, towards a new fixed point. The
scaling behavior about this new fixed point is the topic of the next section.

\section{Kondo Interaction}
\label{section5}

The problem we have studied so far (with only forward electron
scattering off the impurity) is a model problem: we do not expect it
to be observed in the laboratory. The reason is simply that
by inserting a spinful (and chargeless) impurity into a, say, quantum
wire, one necessarily produces a sharp scattering potential, the
core of the interaction being a localized spin exchange. In
contrast, the exclusion of electron back scattering off the impurity,
i.e. the exclusion of {\em large} momentum transfers, requires an
effective scattering potential that varies slowly in space. To have
an experimentally relevant model we must therefore
incorporate the back scattering term ${\cal H}_B$, as
in Eq. (\ref{B}), and study  the {\em full} electron-impurity (Kondo)
interaction ${\cal H}_F + {\cal H}_B$ in the strong-coupling,
low-temperature regime. As we shall see, the insights gained from the
treatment of the forward-scattering problem will turn out to be crucial
in attacking this more difficult problem.

Before plunging into the analysis, we briefly comment upon the recent
work by Furusaki and Nagaosa \cite{FN} who exploit a version of
``poor-man scaling'' to analyze the problem. As for the ordinary Kondo
problem \cite{AYH}, a set of scaling equations is derived perturbatively
in the limit of weak electron-impurity coupling, and tentatively extended
into the strong-coupling regime. This procedure suggests that the coupling
increases indefinitely, implying a ground state where the impurity spin
locks into a singlet with the conduction electrons, causing the chain to
break. A drawback of this method is that the scaling equations are formally
valid only for weak couplings, and may miss out on possible
intermediate-coupling fixed points, as in the two-channel Kondo effect.
However, considering the analogy with the ordinary (one-channel) Kondo problem,
it is quite likely that the result that ensues is in fact valid.
Nonetheless, it is important to make an {\em independent} test of the
result, using a method that is internally consistent.
This is the aim of the following analysis.

\subsection{The non-interacting problem}
\label{section5A}

As a preliminary, let us look at the case of 1D {\em free} electrons
coupled to a localized spin $(S=\onehalf)$ by the Kondo interaction ${\cal
H}_{K} = {\cal H}_{F} + {\cal H}_B$. The total Hamiltonian
${\cal H} = {\cal H}_0 + {\cal H}_K$ is given by
\begin{equation}
{\cal H}_0 = \frac{v_F}{2 \pi} \int dx \biggl[ \no{\psidop{L,\sigma}(x) i
\frac{d}{dx} \psiop{L,\sigma}(x) } - \no{ \psidop{R,\sigma}(x) i
\frac{d}{dx} \psiop{R,\sigma}(x) } \biggl]
\label{H0}
\end{equation}
\begin{equation}
{\cal H}_K = {\lambda}_K \no{ \left[ \psidop{L,\sigma}(0) +
\psidop{R,\sigma}(0)
\right] \onehalf \bsigma{\sigma \mu}  \left[ \psiop{L,\mu}(0) +
\psiop{R,\mu}(0) \right] } \cdot \vS .
\label{HK}
\end{equation}
Noting that only the even-parity part of the electron field couples
to the impurity, it is convenient to pass to a {\em Weyl basis} via
the canonical transformation
\begin{equation}
\psiop{\pm,\sigma}(x) = \frac{1}{\sqrt{2}} \left[ \psiop{L,\sigma}(x) \pm
\psiop{R,\sigma}(-x) \right] .
\end{equation}
{}From this construction it follows that both $\psiop{+,\sigma}$ and
$\psiop{-,\sigma}$ are chiral, left-moving fields with definite
parity $P$ under $x \rightarrow -x$:
\begin{equation}
P \psiop{\pm,\sigma}(x) = \pm \psiop{\pm,\sigma}(x).
\end{equation}
In the
 basis the total Hamiltonian takes the form
\begin{equation}
{\cal H}_0 + {\cal H}_K = \frac{v_F}{2 \pi} \sum_{r = +,-} \int dx
\no{\psidop{r,\sigma}(x) i \frac{d}{dx} \psiop{r,\sigma}(x) }
\ + 2 \lambda \, \no{ \psidop{+,\sigma}(0) \onehalf \bsigma{\sigma \mu}
\psiop{+,\mu}(0) } \cdot \vS. \label{weylH}
\end{equation}
We recognize (\ref{weylH}) as {\em identical} to the Hamiltonian
representing three-dimensional free electrons in two channels (``$+$''
and ``$-$''), coupled to a Kondo impurity in the ``$+$''-channel only.
This leads to a one-channel Kondo fixed point with a $\pi/2$
phase shift of the single-electron levels in the ``$+$'' channel, the
``$-$'' channel being unaffected. (In fact, as pointed out already by
Nozi\`{e}res and Blandin \cite{NB} (see also \cite{ALPC}), the
same conclusion holds
for any channel-anisotropic Kondo interaction: the screening of the
impurity is fully attained by the electrons in the more strongly
coupled channel.)

The low-temperature impurity thermodynamics of the one-channel
problem is that of a local Fermi liquid \cite{Nozieres}. In
particular, the impurity specific heat $C_{imp}$ and magnetic
susceptibility $\chi_{imp}$ scale as
\begin{equation}
C_{imp} \sim  T + O(T^2), \ \ \ \ \ \chi_{imp} \sim T^0 + O(T).
\label{FLscaling}
\end{equation}
{}From (\ref{weylH}), this result also holds for {\em one-dimensional}
free electrons coupled to a spin-$\onehalf$ Kondo impurity.

\subsection{The interacting problem}

The construction above is no longer useful when the electron-electron
interaction is included, that is, when
\begin{eqnarray}
{\cal H}_{int}  =  \frac{1}{2 \pi} \int dx & \biggl\{ &
  \frac{g}{2} \sum_{r,s = L, R}
  \no{ \psidop{r,\sigma}(x) \psiop{r,\sigma}(x) }
\no{ \psidop{s, -\sigma}(x) \psiop{s, -\sigma}(x) } \nonumber
\\
 & + &  \, g \, \no{ \psidop{R,\sigma}(x) \psiop{L,\sigma}(x)
\psidop{L,-\sigma}(x) \psiop{R,-\sigma}(x) } \biggr\} \label{int}
\end{eqnarray}
is added to ${\cal H}_0 + {\cal H}_K$:
The interaction in (\ref{int}) mixes left- and right-moving fields,
and hence becomes non-local in the Weyl basis.

To make progress we must take a less direct route. We shall here exploit the
expectation (see Sec.~\ref{section1} and Ref. \cite{Lrev})
that the full Kondo interaction ${\cal H}_K$ can be described as a
renormalized boundary condition on ${\cal H}^*_{TL}$, analogous to the
case of the ``forward'' interaction ${\cal H}_F$ studied in the
previous sections. This is indeed a well-founded assumption:
In the noninteracting limit ($g = 0$) a
canonical transformation on the even-parity spin current removes
the impurity from the Hamiltonian in the Weyl basis
(cf. Sec.~\ref{section5A}). This, ``by construction''
automatically leads to a change of boundary condition on the critical
bulk theory. Turning on the bulk interaction, this boundary condition
must still be present, although its effect (coded in the new selection rule
for combining conformal towers) may change with a variation of the bulk
coupling. However --- as we have just seen --- when $g \neq 0$, ${\cal H}_K$
cannot be reformulated in terms of spin currents without
violating locality of the electron-electron interaction.
Therefore, we cannot identify the correct selection rule
(boundary condition) by a redefinition of the spin current, as we did for
${\cal H}_F$ in Section~\ref{section3}. In fact, we should not even expect
that the new selection rule is simply related to the old one by a
recombination of conformal towers in the spin sector only.
To the contrary: Since the Kondo interaction ${\cal H}_K$ carries a
charge component in the chiral basis, the charge sector is affected too.
In other words, the selection rule for combining the two charge conformal
towers may also change. This implies that the boundary operators
may be composites of non-trivial operators from the spin {\em and} charge
sectors. The fact that these are described by {\em two distinct conformal
field theories} signals the novel aspect of the problem.

To obtain a sufficiently general framework for this new situation, we
introduce a notation that does not make an implicit relation between the
two charge towers, and denote a general combination of conformal towers by
\begin{equation}
(C_1, D_1; \ C_2, D_2 ; \ j; \ \phi). \nonumber
\end{equation}
Here $(C_i, D_i)$
labels the $U(1)$ channel-$i$ tower ($i=1,2$), while $j$ and $\phi$,
as before, denote the $SU(2)_2$ and Ising towers. The two $U(1)$
towers are now treated as {\em independent}. In the description of ${\cal
H}_{TL}^* + {\cal H}_F$, as well as that of ${\cal
H}_{TL}^*
$ with a trivial boundary, $C_1=C_2 = Q $ and $D_1 =
D_2 = \Delta Q $. These identities are the above-mentioned implicit
relations between the charge towers and may be interpreted as part
of the corresponding selection rule, whereas more general boundary
conditions, associated with other types of impurity interactions,
like ${\cal H}_K$,
may require $C_1 \neq C_2 $ and $D_1 \neq D_2 $. Note that we still
require $C_i \pm  D_i$ to be even for $i=1,2$, so as to preserve the
spectrum {\em within} each conformal tower.

The next step is to find the selection rule, and we first
focus on the effect of ${\cal H}_K$ on the charge sector. As the quantum
numbers $C_i$ and $D_i$ can take any integer values, we assume that any
selection rule for combining the $U(1)$ conformal towers can be expressed by a
linear relation:
\bml
\label{selection_general}
\begin{eqnarray}
 C_1 & = & \alpha \, C_2 + \beta  D_2 + \eta  \label{sg1} \, , \\
 D_1 & = & \gamma \,  C_2 + \, \delta D_2 + \zeta  \, , \label{sg2}
\end{eqnarray}
\eml
with
$\alpha, \beta, ... \zeta $ integers satisfying $\alpha\delta -
\beta\gamma \neq 0$. As we shall see, a few
symmetry constraints severely limit the number of possibilities and
leave us with only two permissible selection rules.
Given these two rules for the $U(1)\times U(1)$ charge sector, we
then consider all possible combinations of conformal towers from
the $SU(2)_2\times$ Ising sector.
Each combination of states corresponds to a (composite) boundary operator, and
those with scaling dimensions $ \geqslant 1$ are candidates for being
the {\em leading-correction-to scaling boundary operator (LCBO)} that
governs the critical
behavior at the ``Kondo fixed point'' in a Luttinger liquid.
As the Kondo interaction (\ref{HK}) breaks the chiral $U(1)$ (as
well as the $SU(2)$) invariance of ${\cal H}_{TL}^*$, $C_i$ and $D_i$
of the {\em LCBO} may now take non-zero values. However, {\em global} $U(1)$
(as well as $SU(2)$) remains a symmetry, imposing
other, weaker, constraints on these quantum numbers.
(The conditions on the quantum numbers in the
$SU(2)_2$ and Ising sectors are as before.)
The list of candidate operators is then restricted by requiring that
any {\em LCBO} respects the symmetry of the original Hamiltonian
(including ${\cal H}_K)$, {\em and} that the Fermi liquid scaling in
(\ref{FLscaling}) is correctly reproduced as $g \rightarrow 0$. (Note
that a selection rule defines a {\em boundary} fixed point, and is valid
for all values of the marginal bulk coupling $g$. Hence, given a
selection rule, Fermi liquid scaling must emerge in the limit $g
\rightarrow 0$.)

With these preliminaries, let us make a first list of candidate
{\em LCBO} scaling dimensions at the ${\cal H}_{TL} + {\cal H}_K$
fixed point. From (\ref{Cdim}), (\ref{SU2dim}), and (\ref{Isingdim}),
treating the two $U(1)$ towers in (\ref{Cdim}) as independent, we have
\begin{equation}
\Delta = \Delta_c + \Delta_s \, , \label{candlist}
\end{equation}
where
\begin{equation}
\Delta_c = \frac{1}{4} \left\{ (q_1)^2 + (q_2)^2 \right\}
\  + N_c \ , \ N_c \in \N ,
\label{calCdim}
\end{equation}
with
\begin{equation}
\label{newqnumbers}
\left\{ \begin{array}{ccccl}
 q_1 & = & \onehalf C_1 \, e^{\theta} & + & \onehalf D_1 \, e^{-\theta} \\
 q_2 & = & \onehalf C_2 \, e^{\theta} & - & \onehalf D_2 \, e^{-\theta} \ ,
\end{array} \right.
\end{equation}
(replacing the former $q^i$ of (\ref{qnumbers})) and
\begin{equation}
\Delta_s = \left\{
\begin{array}{l}
0 \\ \frac{3}{16} \\ \frac{1}{2} \\
\end{array} \right\} +
\left\{
\begin{array}{l}
0 \\ \frac{1}{16} \\ \frac{1}{2} \\
\end{array} \right\} + N_s , \ N_s \in \N . \label{calSdim}
\end{equation}

To cut down the list of possible scaling dimensions in (\ref{candlist}),
we next study the constraints imposed by the symmetries of the model.

\subsection{Symmetries and selection rules}
\label{section5C}

All states in the charge sector are combinations of states from the two
$U(1)$ conformal towers labeled by $(C_1,D_1)$ and $(C_2,D_2)$, or $q_1$ and
$q_2$ for short. Similarly, we label the Kac-Moody primary states of these
conformal towers $\ket{q_1}$ and $\ket{q_2}$, respectively. They
transform under independent $U(1)$-transformations as
\begin{equation}
\ket{q_j} \rightarrow e^{i q_j \phi_j} \ket{q_j}.
\end{equation}
It is convenient to introduce the linear combinations
\begin{equation}
\left\{ \begin{array}{rcccccl}
 q        & \equiv & q_1 + q_2 &
   = & \onehalf (C_1+C_2) e^{\theta} & + & \onehalf (D_1-D_2) e^{-\theta} \\
 \Delta q & \equiv & q_1 - q_2 &
   = & \onehalf (C_1-C_2) e^{\theta} & + & \onehalf (D_1+D_2) e^{-\theta}
 \end{array} \right.
\end{equation}
and
\begin{equation}
\left\{ \begin{array}{rcl}
 \phi        & \equiv & \onehalf (\phi_1 + \phi_2) \\
 \Delta \phi & \equiv & \onehalf (\phi_1 - \phi_2),
 \end{array} \right.
\end{equation}
so that we get a general combination of $U(1)$ transformations as
\begin{equation}
\ket{q_1} \otimes \ket{q_2} \equiv \
\ket{q_1,q_2} \rightarrow e^{i \varphi} \ket{q_1,q_2}
\end{equation}
with
\begin{equation}
\varphi = q_1 \phi_1 + q_2 \phi_2 = q \phi + \Delta q \Delta \phi.
\end{equation}
In terms of these phase factors, global $U(1)$ is generated by $\phi$
and chiral $U(1)$ by $\phi$ and $\Delta \phi$.
Hence, $\ket{q_1,q_2}$ is global $U(1)$ invariant if $q=0$ and chiral
$U(1)$ invariant if $q = \Delta q = 0$. This is consistent with our
previous notion of global and chiral $U(1)$ invariance in terms of $Q$
and $\Delta Q$, because $q = Q e^{\theta}$ and $\Delta q = \Delta Q
e^{-\theta}$ for the (bulk) Luttinger liquid {\em and} forward
scattering problem, where $Q=C_1=C_2$ and $\Delta
Q = D_1 = D_2$.

Consider now the original fixed point theory
${\cal H}_{TL}^*$ without impurity.
Adding ${\cal H}_K$ breaks chiral, but not global, $U(1)$ invariance.
As ${\cal H}_{TL}^*$ satisfies the Luttinger liquid selection rule,
we see that the effect of adding the perturbation ${\cal H}_K$ is to break
$\Delta \phi$ invariance, i.e. to remove the constraint $\Delta q = 0$.
Under renormalization the theory flows to a new fixed point, also
associated with ${\cal H}_{TL}^*$, but with a different selection rule,
where, possibly, the two $U(1)$ towers are decoupled. Whereas $Q$
and $\Delta Q$ can then no longer be used to label the $U(1)$ sector,
$q$ and $\Delta q$ remain well-defined.
Hence, at the new Kondo fixed point, the signature of adding backward
scattering against the impurity is to allow operators with $\Delta q \neq 0$.
The same arguments lead to the requirement $q=0$, in order to preserve
the global $U(1)$ invariance of ${\cal H}_K$.
An {\em LCBO} must therefore not only be compliant with the selection rules
(\ref{selection_general}), but also with
\begin{equation}
\label{globalU(1)inv}
 C_1=-C_2 \ \ \mbox{and} \ \ D_1=D_2.
\end{equation}

Next we consider the effect of a discrete symmetry. Although ${\cal
H}_K$ breaks chiral $U(1)$ symmetry, it is invariant under channel-exchange
${\cal E} : 1 \leftrightarrow 2$, or equivalently $L \leftrightarrow R$ in
(\ref{B}). This translates into invariance under $(q,\Delta q)
\rightarrow (q,- \Delta q)$,
which is equivalent to invariance under
\begin{equation}
\label{Einv}
 {\cal E}: \   C_1 \leftrightarrow C_2 \ \ \mbox{and} \ \ D_1
 \leftrightarrow - D_2.
\end{equation}
Any candidate {\em LCBO} must respect this symmetry. In case $q = \Delta q
=0$, this is trivially fulfilled and $\Delta_c = 0$.
We will return to this special case later and now focus on $\Delta q \neq
0$. Invariance under (\ref{Einv}) then implies that
a {\em LCBO} must be a symmetric combination of operators with opposite
signs of $\Delta q$. To force the coexistence of two such operators,
we have to constrain the selection rules (\ref{selection_general})
to be invariant under (\ref{Einv}), i.e. they must satisfy
\bml
\label{selection_general'}
\begin{eqnarray}
 C_2 & = & \alpha \, C_1 - \beta  D_1 + \eta  \, ,\label{sg1'} \\
 - D_2 & = & \gamma \, C_1 - \, \delta D_1 + \zeta \, .\label{sg2'}
\end{eqnarray}
\eml

If we now require at least one boundary operator, there must be a
solution to (\ref{selection_general}), (\ref{globalU(1)inv}) and
(\ref{selection_general'}). Inserting (\ref{globalU(1)inv})
in (\ref{sg1}) and (\ref{sg1'}) yields $(1+\alpha) C_1 = \beta D_1 + \eta$
and $(1+\alpha) C_1 = \beta D_1 - \eta$, respectively, i.e. $\eta = 0$.
Using (\ref{sg2})
and (\ref{sg2'}), we similarly get $\zeta = 0$. We may therefore conclude
that $\eta = \zeta = 0$ is a necessary condition for the selection rules.

By demanding full consistency between (\ref{selection_general}) and
(\ref{selection_general'}) we may further reduce the list of possible
selection rules: For instance, combining
(\ref{sg1}) and (\ref{sg1'}) requires $(1-\alpha^2) C_1 = - \alpha \beta D_1
+ \beta D_2$. Let us first consider the case $\alpha^2 \neq 1$.  Then
$\alpha \beta = 0$, as otherwise $C_1$ would be a function of $D_1$.
(Remember that selection rules only give relations between conformal towers
and should not pose constraints within.) $\beta = 0$ implies $C_1 =
0$, which is not allowed by the same reason. The other possibility,
$\alpha=0$, implies $C_1 = \beta D_2$. However, as $C_1$ may be any
integer, we can only allow $\beta = \pm 1$. Inserting this relation in
(\ref{sg2'}), yields $(\beta^{-1}+\gamma)C_1 = \delta D_1$. To avoid
constraints within a conformal tower, we must then require
$\beta^{-1}+\gamma = \delta = 0$. Hence, the only solution for
$\alpha^2 \neq 1$
is $\alpha = \delta = 0$ and $\beta = -\gamma = \pm 1$.
The next case, $\alpha = 1$, implies $\beta D_1 = \beta D_2$. If we
assume $\beta \neq 0$, then $D_1 = D_2$, which gives $C_1 = C_2 + \beta
D_1$ using (\ref{sg1}). But then $\beta = 0$, as $C_1$ cannot be a function
of $D_1$, i.e. $\beta \neq 0$ leads to a contradiction. The only
possibility for $\alpha =1$ therefore is to require $\beta = 0$, which
does not lead to a contradiction. The same result holds for $\alpha = -1$.

Analogous treatment of (\ref{sg2}) and (\ref{sg2'}) implies that either
$\alpha = \delta = 0$ and $\beta = -\gamma = \pm 1$
or else $\delta = \pm 1$ and $\gamma = 0$. Hence, in total there are only six
possible selection rules in the charge sector:
\bml
\label{sr1-4}
\begin{eqnarray}
 C_1 = C_2  & \ \mbox{and} \ & D_1 = D_2  \label{sr1} \\
 C_1 = C_2 & \ \mbox{and} \ & D_1 = -D_2  \label{sr2} \\
 C_1 = -C_2  & \ \mbox{and} \ & D_1 = D_2 \label{sr3} \\
 C_1 = -C_2 & \ \mbox{and} \ & D_1 = -D_2 \label{sr4}
\end{eqnarray}
\eml
and
\bml
\label{sr5-6}
\begin{eqnarray}
 C_1 = D_2  & \ \mbox{and} \ & D_1 = -C_2 \label{sr5} \\
 C_1 = -D_2 & \ \mbox{and} \ & D_1 = C_2  \label{sr6}.
\end{eqnarray}
\eml

The first selection rule (\ref{sr1}) is the Luttinger liquid selection rule
that we start off with before we include Kondo scattering off the impurity.
The effect of including ${\cal H}_K$ is to move us to a new fixed point, which
may be described by any of the above six selection rules. In analogy to
changing fixed point in the case of forward scattering off the impurity, or any
other quantum impurity problem, we shall call such a transformation a {\em
fusion} in the charge sector. It is a prescription for how the conformal
towers are recombined when we change fixed points. From (\ref{sr1-4}) and
(\ref{sr5-6}) it follows that there are six possible fusion rules that
can be applied to the Luttinger liquid selection rule, and one
of these should correspond to adding ${\cal
H}_K$ to the fixed-point Hamiltonian. For instance, the fusion rule
$(C_2,D_2) \rightarrow (-C_2,D_2)$ changes (\ref{sr1}) to (\ref{sr3}) and
$(C_2,D_2) \rightarrow (D_2, -C_2)$ changes (\ref{sr1}) to
(\ref{sr5}). Applying the correct fusion rule {\em once} to (\ref{sr1}) gives
the new Kondo fixed point, and the selection rule can be used to extract
the finite-size energy spectrum. Furthermore, we expect that applying the
same fusion rule {\em twice} should
give us the selection rule that determines the boundary scaling dimensions.
It is easy to check that any fusion rule that takes (\ref{sr1}) to
any of the selection rules in (\ref{sr1-4}) gives (\ref{sr1}) back after
double fusion, whereas
the fusion rules that take (\ref{sr1}) to any of (\ref{sr5-6}) give
(\ref{sr4}) after double fusion. We therefore conclude that the only
possible selection rules for the boundary scaling dimensions are
(\ref{sr1}) and (\ref{sr4}).

We may now apply the symmetry constraint (\ref{globalU(1)inv}) to extract
a ``short list'' of boundary scaling dimensions from the charge sector.
The first rule, (\ref{sr1}), together
with (\ref{globalU(1)inv}), requires an {\em LCBO} to have $C_1=C_2=0$ and
$D_1=D_2$ an even integer. Using (\ref{calCdim}) and (\ref{newqnumbers}),
\begin{equation}
\label{deltac1}
\Delta_c = \onehalf p^2  e^{-2\theta} + N_c ,
  \ \  p, N_c \in \N.
\end{equation}
Similarly, the second selection rule, (\ref{sr4}), combined with
(\ref{globalU(1)inv}), requires $D_1 = D_2 = 0$ and $C_1 =
-C_2$ an even integer, i.e.
\begin{equation}
\label{deltac2}
\Delta_c = \onehalf p^2  e^{2\theta} + N_c ,
  \ \ p, N_c \in \N.
\end{equation}

The full boundary dimensions are obtained by coupling the $SU(2)_2$
and Ising conformal towers to the pairs of $U(1)$ towers in (\ref{sr1})
and (\ref{sr4}), respectively. Starting with the $SU(2)_2$ sector, the
$j=\onehalf$ tower is expelled by global $SU(2)_2$ invariance: Spin
rotational invariance of the Hamiltonian ${\cal H}_{TL} + {\cal H}_K$
implies that
any {\em LCBO} must transform as a spin singlet, which, however, is missing
from the $j=\onehalf$ tower. Turning to the $j=1$ tower, the primary operator
$\vphi$ is expelled by the same reason. The lowest-dimension
$SU(2)_2$ singlet operator from this tower
is $\vJ_{-1} \cdot \vphi$. However, this is the same operator
that drives critical scaling in the forward scattering problem. In
particular, it produces a diverging impurity susceptibility as $T
\rightarrow 0$ (cf. (\ref{CHIimp})), in conflict with the known Fermi
liquid scaling (\ref{FLscaling}) in the $g \rightarrow 0$ limit of the present
problem.\footnote{This conclusion still holds in the presence of
non-trivial operator factors from the $U(1)$ and/or Ising sectors.
Whatever their dimensions, these can produce only $T-$independent
constants in the scaling of the susceptibility with
temperature (see Sec.~\ref{section5E}).} Assigning
evenness to the $j=1$ conformal tower under channel exchange (as in the
forward scattering problem) implies that $\vJ_{-1} \cdot \vphi$ is
allowed by symmetry, and hence any selection rule must suppress this
tower. The reverse assignment of parity under channel exchange instead
implies that the $j=1$ tower is suppressed by symmetry.
Summarizing,
the only possible contributions to an {\em LCBO} from the $SU(2)_2$ sector,
consistent with established results for $g=0$, are the identity operator
and its descendants.
We are thus left with the problem of gluing together the pairs of $U(1)$ towers
in (\ref{sr1})
and (\ref{sr4}) with the towers in the Ising sector.

Let us start with (\ref{sr1}). Putting $D_1 = D_2 = 0$, the resulting
identity towers can be combined only with the identity tower in the Ising
sector (together with that of $SU(2)_2$). This is so, since the presence
of a $\phi = \sigma $ or $\epsilon$ tower would lead to a relevant
boundary operator: the primary operators $\sigma$ and $\epsilon$ both
have dimensions $< 1$. However, at $g=0$ the fixed point is known to be
stable (cf. Sec.~\ref{section5A}), excluding the presence of a relevant
operator. In fact, this conclusion may be extended to $g \neq 0$:
To remain at an unstable fixed point (that is, to maintain criticality)
requires fine tuning of some parameter in the bare Hamiltonian.
As $g$ is the only tunable parameter in ${\cal H}^*_{TL}$ (with a
renormalized boundary condition replacing $H_K$), an unstable fixed point
would imply non-criticality for all values of $g \neq 0$. In other words,
the total scaling dimension $\Delta=\Delta_c + \Delta_s$ of any boundary
operator must be $> 1$. It is here important to stress
that --- by the {\em raison d'etre} of renormalization --- {\em any}
boundary operator allowed by symmetry will also appear at the fixed point.
Therefore, one cannot argue that an {\em LCBO} with $\Delta >1$ can be
obtained by forming descendants in the $\phi = \sigma $ or $\epsilon$ towers.
If any of these towers were present, the corresponding primary operators of
dimension $\Delta =\frac{1}{16}$ and $\onehalf$, respectively, would be present
as well, implying an unstable fixed point.

The lowest-dimension operators
emerging from the identity towers are the first Kac-Moody descendants in the
$U(1)$ sectors, with $\Delta = 1$: The marginal boundary operators
${\cal O}^{1,2}(w) = j^{1,2}_L(w)$ always appear in the charge sector, as the
particle-hole symmetry of the original lattice model in (\ref{Hubbard}) is
broken away from half-filling.\footnote{Although the Hubbard model
(\ref{Hubbard}) is not invariant under charge conjugation (particle-hole
transformation) $\cop{n,\sigma} \rightarrow (-1)^n
\epsilon^{\phdagger}_{\sigma \mu} \cdop{n, \mu}$ off half-filling, its
bosonized counterpart, the Tomonaga-Luttinger model (\ref{TL}), is invariant
under the corresponding transformation $\psiop{r,\sigma}(x) \rightarrow
\epsilon^{\phdagger}_{\sigma \mu} \psidop{r,\mu}(x)$.
The reason for this is that the normal ordering of the
latter subtracts the symmetry-breaking terms. We may remove this
accidental symmetry by explicitly introducing operators that break charge
conjugation.} Upholding particle-hole symmetry, the lowest dimension
operators would instead be the second {\em Virasoro descendants} $L_{-2} \1$
in respective sectors, of dimensions $\Delta = 2$.
The next choice of $U(1)$ quantum numbers, $D_1 = D_2 = 2 \ (p=1$ in
(\ref{deltac1})), leads to a relevant boundary operator for any combination
of Ising towers, and is therefore not allowed.
In contrast, $D_1 = D_2 \geqslant 4 \ (p \geqslant 2$ in (\ref{deltac1}))
yields
operators with $\Delta \geqslant 1$ when combined with any Ising tower.
Summarizing, the possible couplings of Ising conformal towers to the
$U(1)$ towers selected by (\ref{sr1}) yield the following candidate
{\em LCBO} dimensions:
\begin{equation}
\Delta_{LCBO} = 1, \ \onehalf p^2 e^{-2\theta} +
\left\{ \mbox{$0, \ \frac{1}{16}, \ \frac{1}{2}$} \right\},
 \ \ \ p \in \N + 2.
\label{candI}
\end{equation}
Turning to the second selection rule for the $U(1) \times U(1)$ sector,
(\ref{sr4}), employing the same reasoning as above, one finds a
second class of possible {\em LCBO} dimensions:
\begin{equation}
\Delta_{LCBO} = 1, \ \onehalf e^{2\theta} + \onehalf, \
\onehalf p^2 e^{2\theta} +
\left\{ \mbox{$0, \ \frac{1}{16}, \ \frac{1}{2}$} \right\},
\ \ \ p \in \N + 2. \label{candII}
\end{equation}

Before exploring the critical behavior implied by the various dimensions
in (\ref{candI}) and (\ref{candII}), two comments may be in order. First,
note that no scaling dimensions of descendant operators --- other than
those of the identity --- appear in (\ref{candI}) or (\ref{candII}).
This is so, since an {\em LCBO} is the
(composite) boundary operator with {\em lowest} dimension $\geqslant 1$,
given a particular combination of conformal towers. Only if one, or
several, of the non-trivial primary operators are expelled by symmetry can a
descendant operator enter the stage (as in the forward scattering problem).
This is not the case here. Secondly, the appearance of the
$\epsilon$ conformal tower requires the reverse assignment of
parity under channel exchange, as compared to the forward scattering
problem. That is, a consistent representation now forces the $\epsilon$
tower to be invariant under channel exchange, while  $\vphi$ and its
descendants (already suppressed by the selection rules) change sign.
(Note that there is no contradiction with the forward scattering case, as the
argument in Sec.~\ref{section4B}
for the other assignment of parity no longer applies.)

\subsection{Impurity specific heat}
\label{section5D}

Analogous to the forward scattering problem, an effective scaling
Hamiltonian ${\cal H}$ is obtained by adding a
boundary term to the fixed point theory of ${\cal H}_{TL} = {\cal H}_0 +
{\cal H}_{int}$:
\begin{equation}
{\cal H} = {\cal H}^*_{TL} + \lambda_I
{\cal O}(0), \label{scalingH}
\end{equation}
where ${\cal O}(0)$ is an $LCBO$ with conjugate scaling field $\lambda_I$.
By mapping the half-plane $\Cplus = \{\im z > 0 \}$ for zero temperature
onto the finite-$T$ geometry (Fig.~\ref{fig:finiteT}), $ \Gammaplus =
\{w= v \tau + ix = (v\beta/\pi) \, \mbox{arctan}(z)\}$ (with $v=v_c (v_s)$
for an $LCBO$ from the charge (spin) sector\footnote{The
finite-temperature geometry is {\em defined} in terms of $\tau$ and $x$ by
$\Gammaplus = \{ -\beta/2 \leqslant \tau \leqslant \beta/2, \
x \geqslant 0 \}$ with periodic boundary condition in $\tau$.
When $v_c \neq v_s$, we {\em represent} it in complex notation
by cylinders of different circumferences
for charge and spin degrees of freedom, respectively. The only reason for
introducing this complex notation is to simplify the derivation of
correlation functions, whereas the calculation of the free energy is
preferably done in terms of $\tau$ and $x$.}), the partition function in zero
magnetic field is written
\begin{equation}
e^{-\beta F(\beta,\lambda_I)} = e^{-\beta F(\beta,0)}
\cf{ e^{ \lambda_I \int_{-\beta/2}^{\beta/2} d\tau
{\tilde {\cal O}}(\tau, 0) } }_T \, ,
\end{equation}
so that $\delta f_{imp}(\beta,\lambda_I) \equiv f_{imp}(\beta,\lambda_I)
- f_{imp}(\beta,0)$ satisfies
\begin{equation}
e^{-\beta \delta f_{imp}(\beta,\lambda_I)} = \ \cf{ e^{ \lambda_I
\int_{-\beta/2}^{\beta/2} d\tau{\tilde {\cal O}}(\tau, 0) } }_T.
\label{fimp}
\end{equation}
We have here used the decomposition in (\ref{Free})
and passed to a Lagrangian formalism, ``tilde'' and $\cf{\ }_T$
referring to $\Gammaplus$.
By a linked cluster expansion,
\begin{equation}
\delta f_{imp} = -\frac{\lambda_I}{\beta} \int_{-\beta/2}^{\beta/2} d\tau
\cf{{\tilde {\cal O}}(\tau,0)}_T  -\frac{\lambda_I^2}{2\beta}
\int \! \! \int_{-\beta/2}^{\beta/2} d\tau_1 \, d\tau_2
\cf{{\tilde {\cal O}}(\tau_1, 0) {\tilde {\cal O}}(\tau_2, 0)}_{T, c} +
O(\lambda_I^3),
\label{linkc}
\end{equation}
with $\cf{\ }_{T,c}$ denoting a cumulant in $\Gammaplus$.

Here two cases must be distinguished: (i) ${\cal O}$ is
Virasoro descendant of $\1$. Then $\cf{{\tilde {\cal O}}(\tau,0)}_T$
may be non-zero and hence the leading contribution to $\delta f_{imp}$
will be linear in $\lambda_I$. (ii) In any sector of
the theory, ${\cal O}$ is Virasoro primary or a Virasoro descendant
of an operator other than $\1$. Then $\cf{{\tilde {\cal O}}(\tau,0)}_T = 0$
and $\delta f_{imp}$ is quadratic in $\lambda_I$.
To see how this comes about, consider a chiral (say, left-moving)
Virasoro primary operator ${\cal A}(z) \neq \1$ with dimension
$\Delta \ (\neq 0)$ in the half-plane $\Cplus$.
The expectation value of a chiral operator in a halfplane is the same as
in the full plane, as translational invariance in one direction implies this
in all directions. (Uniqueness of analytic functions implies that the
expectation value is constant everywhere.) The scale transformation
$z \rightarrow z/a$, where we later choose $a=z$, implies
\begin{equation}
 \cf{{\cal A}(z)} = z^{-\Delta}\cf{{\cal A}(1)} ,
\end{equation}
and we conclude that $\cf{{\cal A}(z)}=0$.
(Note that the argument crucially depends on ${\cal A}$ being chiral:
a non-chiral operator may pick up a non-zero expectation value in the
presence of a boundary.) Any Virasoro descendant operator
$L_{-n_1} L_{-n_2} \dots L_{-n_M} {\cal A}(z)$ ($n_j > 0$) has a vanishing
expectation value as well, since $ \cf{L_{-n_1} \ldots L_{-n_M} {\cal
 A}(z)} = {\cal L}_{-n_1} \ldots {\cal L}_{-n_M}\cf{A(z)}= 0,
\ {\cal L}_{-n_j} \ (j = 1, \dots ,n_M) $
being differential operators \cite{BPZ}.
Now map $\Cplus$ onto $\Gammaplus$.  Again, by conformal invariance,
$\cf{{\tilde {\cal A}}(w)}_T = (\frac{dw}{dz})^{-\Delta} \cf{A(z)} = 0$.
The transformation law for a Virasoro descendant is more complicated and
relates one descendant in $\Gammaplus$ with a sum of operators from
the same Virasoro tower in $\Cplus$. However, as the expectation
values of these are zero, it follows that the expectation value of a
Virasoro descendant of ${\cal A}$ vanishes in $\Gammaplus$ as well.
In contrast to this, descendants of the unit operator $\1$ may acquire
non-zero expectation values in $\Gammaplus$. As an example, the
energy-momentum tensor is a descendant of $\1$ and satisfies
$\cf{{\tilde T}(w)}_T = \frac{c}{12}\{z, w \}$, with $c$ the conformal
anomaly number and $\{z, w \} = \frac{d^3z/dw^3}{dz/dw}
- \frac{3}{2} \left(\frac{d^2z/dw^2}{dz/dw} \right)^2$ the Schwarzian
derivative of the map $\Cplus \rightarrow \Gammaplus$.

Let us first study case (i) where ${\cal O}$ is a descendant of $\1$.
Since $L_{-1} \1 = d\1/dz = 0$, the energy momentum tensor
$T(z) \equiv L_{-2} \1$ is the leading Virasoro descendant of $\1$
and has $\Delta = 2$.
In the present problem, each sector contributes its own energy momentum
tensor, $T_1(z) = \frac{1}{4} \no{j^1_L(z) j^1_L(z)}$,
$T_2(z) = \frac{1}{4} \no{j^2_L(z) j^2_L(z)}$,
$T_3(z) = \frac{1}{4} \no{\vJ(z) \cdot \vJ(z)}$ and $T_4(z) = T_{Ising}(z)$,
all with $\Delta = 2$ and satisfying $\cf{T_j(z)} = 0$ in $\Cplus$.
In other words, there are {\em four} degenerate {\em LCBOs} for this case.
Passing to $\Gammaplus$, their contribution to the impurity
specific heat is given by
\begin{equation}
\delta f_{imp} = -\sum_{j=1}^{4} \frac{\lambda_j}{\beta} \int_{-\beta
/2}^{\beta /2} d\tau \cf{\tilde{T}_j(\tau,0)}_T, \label{fimp(i)}
\end{equation}
with
\begin{equation}
\cf{\tilde{T}_j(w)}_T = \frac{c_j}{12}\{z, w \} =
\frac{c_j}{6}(\frac{\pi}{v_j\beta})^{2},
\label{Schwarz}
\end{equation}
where $c_1 = c_2 = 1$, $c_3 = \frac{3}{2}$ and $c_4 = \onehalf$ are the
conformal anomaly numbers of the different sectors, and $v_1 = v_2 = v_c$ and
$v_3 = v_4 = v_s$ are the corresponding velocities.
Inserting (\ref{Schwarz}) into (\ref{fimp(i)}), integrating, and
summing over the sectors, it follows that
\begin{equation}
\delta f_{imp} =  -\frac{\pi^2}{6}\sum_{j=1}^4\frac{\lambda_j c_j}{v^2_j} T^2,
\label{(i)result}
\end{equation}
producing a linear specific heat
\begin{equation}
C_{imp} = -T \frac{\partial^2 f_{imp}}{\partial T^2} =
\frac{\pi^2}{3} \sum_{j=1}^4 \frac{\lambda_j c_j}{v^2_j} T,
  \label{Cimp(i)}
\end{equation}
where we have used (\ref{Cimp_crit}).
This is the dominant contribution to $C_{imp}$ that is linear in the
scaling fields: higher-order descendants of $\1$ produce higher
powers in temperature.

We now turn to the more interesting case (ii) where the candidate {\em
LCBOs} are {\em not} Virasoro descendants of the unit operator. The
one-point function in (\ref{linkc}) then vanishes, and we are left with
\begin{equation}
\delta f_{imp} = -\frac{\lambda_I^2}{2\beta}
\int \! \! \int_{-\beta/2}^{\beta/2} d\tau_1 \, d\tau_2
\cf{{\tilde {\cal O}}(\tau_1, 0) {\tilde {\cal O}}(\tau_2, 0)}_{T}
+ O({\lambda}_I^3).
\label{second}
\end{equation}
According to our symmetry analysis, any {\em LCBO} that is {\em not} a
descendant of $\1$ must be a Virasoro primary operator.\footnote{All
Kac-Moody primary operators allowed by symmetry, as well as the first
Kac-Moody descendants
in the charge sector, are Virasoro primary.} It is therefore
sufficient to consider the case when ${\cal O}$ is Virasoro primary, for
which the two-point function in
$\Cplus$ takes the familiar form
\begin{equation}
\cf{{\cal O}(z_1) {\cal O}(z_2)} = \frac{A}{(z_1-z_2)^{2\Delta}}.
\label{twopoint}
\end{equation}
Here $\Delta \ (\neq 0)$ is the scaling dimension of ${\cal O}$, and $A$ is a
normalization constant (to be determined). Using $z \rightarrow w = v\tau +
ix = (v\beta/\pi) \, \arctan z$ and the transformation rule for a
Virasoro primary operator, we can get
$\cf{\tilde{{\cal O}}(w_1)\tilde{{\cal O}}(w_2)}_T$.
However, as ${\cal O}$ may be composed of operators from sectors with
different velocities $v$, we must perform the transformation in
each sector independently. The expression simplifies somewhat on the
boundary $x=0$ and becomes
\begin{equation}
\label{twopoint_T}
\cf{\tilde{{\cal O}}(\tau_1,0)\tilde{{\cal O}}(\tau_2,0)}_T \ = \
\frac{A}{(v_c^{\Delta_c} v_s^{\Delta_s})^2
|\frac{\beta}{\pi}\mbox{sin}(\frac{\pi}{\beta}(\tau_1 - \tau_2))|^{2\Delta}} \
{}.
\end{equation}
where $\Delta = \Delta_c + \Delta_s$, and the subscripts
refer to the charge and spin sectors, respectively.
The integrand (\ref{twopoint_T}) of (\ref{second}) is even and periodic,
with period $\beta$, and can be replaced by a single integral over
$\tau = \tau_1 - \tau_2$. Putting $u \equiv \mbox{tan}(\pi/\beta)$, and
inserting a short-time cutoff $\tau_0 = \epsilon \beta/\pi$,
yields the expression
\begin{equation}
\delta f_{imp} = - \frac{\lambda_I^2 A}{(v_c^{\Delta_c} v_s^{\Delta_s})^2 }
(\frac{\pi}{\beta})^{2\Delta-1}
\int_{\tan\epsilon}^{\infty} du \frac{(1+u^2)^{\Delta-1}}{u^{2\Delta}} \ .
\label{fimp-u}
\end{equation}

Let us first study the case when $\Delta \geqslant 1$ is an
integer. We can then make the expansion $(1+u^2)^{\Delta-1} = 1 +
(\Delta-1)u^2 + \ldots + u^{2(\Delta-1)}$, which yields
\begin{equation}
I \equiv \int_{\tan\epsilon}^{\infty} du
\frac{(1+u^2)^{\Delta-1}}{u^{2\Delta}} =
\frac{1}{2\Delta-1} \frac{1}{(\tan\epsilon)^{2\Delta-1}}\left(1+
\frac{\Delta-1}{2\Delta-3}(2\Delta-1)\tan^2 \! \epsilon +
O(\epsilon^4) \right) .
\label{interim}
\end{equation}
In the limit of small $\epsilon$, i.e. $\tau_0 T \rightarrow 0$,
equations (\ref{fimp-u}) and (\ref{interim})
give
\begin{equation}
C_{imp} = -T \frac{\partial^2 f_{imp}}{\partial T^2}
= \frac{\lambda_I^2 A}{(v_c^{\Delta_c} v_s^{\Delta_s})^2 }
 \frac{2\Delta}{3(2\Delta-3)} \pi^2 \tau_0^{3-2\Delta}
T \left( 1 + O [(\tau_0 T)^2] \right) . \label{integerC}
\end{equation}

The case of non-integer dimension $\Delta>1$ requires a more lengthy
analysis. One partial integration of $I$ yields
\begin{equation}
I = \frac{1}{2\Delta-1}(\frac{\beta}{\pi \tau_0})^{2\Delta-1}\left(1 +
\frac{\Delta-2}{3}
(\frac{\pi \tau_0 }{\beta})^2 + O [(\frac{\pi \tau_0}{\beta})^4] \right)
+ \frac{2(\Delta-1)}{2\Delta-1} I_1
\label{partI}
\end{equation}
with $I_1$ defined by
\begin{equation}
I_j = \int_{\tan\epsilon}^{\infty}du \frac{(1+u^2)^{\Delta - j
-1}}{u^{2(\Delta - j)}} \ , \ \ \ j \in \N . \label{Ij}
\end{equation}
When $\Delta < \frac{3}{2}$, $I_1$ is finite as $\epsilon \rightarrow 0$,
while $\Delta = \frac{3}{2}$ produces a logarithmic singularity:
\begin{equation}
I_1 \rightarrow \left\{
\begin{array}{ll}
\frac{1}{2} B(\frac{3}{2}-\Delta, \frac{1}{2}), & \Delta < \frac{3}{2} \\
\ln \frac{1}{\tau_0 T} + \mbox{constant}, & \Delta = \frac{3}{2} \\
\end{array} \right. \ \ \ \ \ \mbox{as} \ \tau_0 T \rightarrow 0
\label{3/2}
\end{equation}
with $B(p,q) = \Gamma(p)\Gamma(q)/\Gamma(p+q)$ the Beta function.

For $\Delta > \frac{3}{2}$, $I_1$ diverges algebraically with the cutoff,
and we must perform a second partial integration to identify the rate of
divergence. This leads to
\begin{equation}
I_1 = \frac{1}{2\Delta -3} (\frac{\beta}{\pi \tau_0})^{2\Delta -3}
\left( 1 + O[(\frac{\pi \tau_0}{\beta})^2] \right) +
\frac{2(\Delta -2)}{2\Delta -3}I_2
\label{I2}
\end{equation}
with $I_2$ defined as in (\ref{Ij}).
In the interval $\frac{3}{2} < \Delta < \frac{5}{2}$, $I_2$ is finite as
$\epsilon \rightarrow 0$, while for $\Delta = \frac{5}{2}$ one gets the same
logarithmic singularity as for $\Delta = \frac{3}{2}$, i.e.
$I_2(\Delta = \frac{5}{2}) = I_1(\Delta =\frac{3}{2})$. The procedure is
now iterated, $I_{j+1}$ being the remainder from partially integrating
$I_{j}$. However, the terms in $T$ thus generated are always {\em subleading}
compared to those coming from the partial integrations of $I$ and $I_1$ in
(\ref{partI}) and (\ref{3/2}), respectively. Thus, collecting the results,
we obtain for the impurity specific heat
\begin{equation}
C_{imp} =
\frac{\lambda_I^2 A}{(v_c^{\Delta_c} v_s^{\Delta_s})^2 } \times
 \left\{ \begin{array}{ll}
   \frac{2\Delta}{3(2\Delta -3)} \pi^2 \tau_0^{3-2\Delta} T + \ldots
      & \Delta = 1 \ \mbox{or} \ \Delta > \frac{3}{2} \\
   2(\Delta-1)^2 \pi^{2\Delta -1} B(\frac{3}{2}-\Delta, \frac{1}{2})
      T^{2\Delta-2} + \ldots  & 1<\Delta <\frac{3}{2}  \\
   \pi^2 T \ln (\frac{1}{\tau_0 T}) + \ldots & \Delta = \frac{3}{2}
\end{array} \right . \label{Cimp(ii)}
\end{equation}
where ``$\ldots$'' denotes subleading corrections.

Combining (\ref{Cimp(i)}) and (\ref{Cimp(ii)}) with the result from the
symmetry analysis, (\ref{candI}) and (\ref{candII}), we find that there
are only two distinct possibilities for critical scaling:
\begin{equation}
\mbox{(i)} \ \ C_{imp} = O(T) \, , \label{nr1}
\end{equation}
and
\begin{equation}
\mbox{(ii)} \ \  C_{imp} =
\frac{\lambda_I^2 A \pi^{1/K_{\rho}}}{2 v_s v_c^{1/K_\rho}}
(\mbox{$\frac{1}{K_{\rho}}$} - 1)^2 B(1-\mbox{$\frac{1}{2K_{\rho}}$},
\onehalf) \, T^{\frac{1}{K_{\rho}} - 1} + O(T) , \label{nr2}
\end{equation}
with $K_{\rho} \equiv e^{-2\theta} = (1+2g/v_F)^{-1/2}$ the Luttinger
liquid charge parameter.
The first case, (\ref{nr1}), is implied when the {\em LCBO} carries dimension
$\Delta = 1$ or $ \Delta > \frac{3}{2}$. In contrast, the leading term in
the second case, (\ref{nr2}), is driven by a composite {\em LCBO} of
dimension $\Delta = \Delta_c + \Delta_s = \onehalf(e^{2\theta} +
1)$, corresponding to the second entry in (\ref{candII}).\footnote{Since we
assume the interaction to be weak $(g \ll 1)$, we take $
 \onehalf(e^{2\theta} + 1) < 3/2$.}
In terms of quantum numbers $(C_1, D_1; \ C_2, D_2; \ j; \ \phi)$,
the operator is given by the sum of
\begin{equation}
\label{LCBO}
(2, 0; \ -2, 0; \ 0; \ \epsilon) \ \ \mbox{and} \ \
(-2, 0; \ 2, 0; \ 0; \ \epsilon).
\end{equation}
Hence, the charge sector contributes a channel-symmetric combination of
primary operators with $q=0$ and $\Delta q = \pm 2 e^\theta$ (cf. discussion
after (\ref{Einv})), which can be explicitly expressed in terms of vertex
operators of free boson fields. The contribution to the scaling dimension
is $\Delta_c = \onehalf e^{2\theta} = 1/2K_\rho$.
Only by combining these operators with the Ising energy density $\epsilon$
does one obtain a boundary operator of dimension $\Delta \geqslant 1$, as
required for an {\em LCBO}.
The linear term in (\ref{nr2}) comes from subleading terms generated by
the same operator, as well as from leading terms due to the marginal
operators $j^1(z)$ and $j^2(z)$. As we have seen, the latter operators are
always present, in case (i) as well, due to the breaking of particle-hole
symmetry.

The specific heat in (\ref{nr2}) exhibits the same anomalous scaling in
temperature as found by Furusaki and Nagaosa \cite{FN}. Also, the way the
anomalous term vanishes as $K_{\rho} \rightarrow 1 \ (g \rightarrow 0)$
is identical to that obtained in \cite{FN}.\footnote{The charge parameter
is defined as $K_{\rho} \equiv [(1-g_2/\pi v_F)/(1+g_2/\pi v_F)]^{1/2}$
in \cite{FN}. For small $g_2$, and switching from the conventional
normalization of Fermi fields, $\{\psiop{r,\sigma}(x), \psidop{s,\mu}(y) \} =
\delta_{r s}  \delta_{\sigma \mu}  \delta(x-y)$, to our normalization in
(\ref{psianticomm}), this is the same as our definition of $K_{\rho}$ with
$g_2 = \pi g$.}
Although more work is needed to firmly establish which of the two cases
applies, (\ref{nr1}) (Fermi liquid) or (\ref{nr2}) (non-Fermi liquid),
the second, non-Fermi liquid case is clearly
favored considering its emergence in an independent analysis. However, a
caveat is advisable. In a recent study, Schiller and Ingersent treats a
simplified model of a magnetic impurity in a ``reduced'' Luttinger liquid,
composed of right-moving, spin-up electrons and left-moving
spin-down electrons \cite{SI}. This problem, with only two branches of
electrons (compared to the four branches of the full problem: two
chiralities with two spin projections each) is mapped exactly onto the
Kondo effect in a Fermi liquid, with $C_{imp}$ as in (\ref{nr1}).
Although the relevance of this simplified model to the Kondo effect in a
full Luttinger liquid remains unclear to us, the result may be
taken - as argued by the authors in \cite{SI} - to give some indirect
support to a Fermi liquid scenario. In any event, it is reassuring
that our exact analysis gives room {\em only} to those two scenarios
that have been conjectured in the literature.

Before closing this section, let us point out that the impurity specific
heat (\ref{Cimp}) for the forward-scattering problem follows from the
third case in (\ref{Cimp(ii)}) by inserting $\Delta_c = 0$, $\Delta_s =
\frac{3}{2}$, and the value of the normalization constant for this case,
$A = \cf{\vJ_{-1} \cdot \vphi | \vJ_{-1} \cdot \vphi} = 9$.\footnote{Note
that $\cf{\vJ_{-1} \cdot \vphi | \vJ_{-1} \cdot \vphi} =
\bra{\phi^a} J^a_{1}J^b_{-1} \ket{\phi^b} = \bra{\phi^a} [J^a_{1} ,
J^b_{-1}] \ket{\phi^b}$, where the second identity follows from
$\ket{\phi^b}$ being a Kac-Moody primary state.
By exploiting the $SU(2)_2$ Kac-Moody algebra, and the fact that $\vphi$
transforms as a spin-1 object under global $SU(2)$ (generated by
$\vJ_{0}$), one straightforwardly arrives at the result, $A=9$.} The
amplitude in (\ref{nr2}) can similarly be calculated
by evaluating the norm of the corresponding {\em LCBO}.
However, as we shall find in the next section, the favored case, with the
{\em LCBO} in (\ref{LCBO}), does not lead to a universal Wilson ratio when
$g \neq 0$. For this reason we here leave the amplitudes undetermined.

\subsection{Impurity susceptibility}
\label{section5E}

In the presence of a magnetic field, the partition function takes the form
\begin{equation}
e^{-\beta F(\beta, h, \lambda_I)} = e^{-\beta F(T,0,0)}
\cf{ e^{ \int_{-\beta/2}^{\beta/2} \left[ \lambda_I d\tau
{\tilde {\cal O}}(\tau, 0) + \frac{h}{2\pi} \int_{-\infty}^{\infty} dx
{\tilde J}^z (\tau, x) \right] } }_T . \label{h-partition}
\end{equation}
The shift of the magnetic susceptibility due to the impurity,
\begin{equation}
\chi_{imp} = - \frac{\partial^2 f_{imp}}{\partial h^2}\Big|_{h=0} \, ,
\end{equation}
may thus be expanded to second order in $\lambda_I$ as
\begin{eqnarray}
\chi_{imp} & =  & \frac{\lambda_I}{4\pi^2 \beta}
\int \! \! \int_{-\infty}^{\infty} dx_1 \, dx_2
\int \! \! \ldots \! \! \int_{-\beta/2}^{\beta/2} d\tau_1 \! \ldots d\tau_3
\cf{\tilde{J}^z(\tau_1, x_1) \tilde{J}^z(\tau_2,x_2)
{\tilde {\cal O}}(\tau_3, 0)}_{T, c}
\nonumber \\
& + & \frac{\lambda_I^2}{8\pi^2 \beta}
\int \! \! \int_{-\infty}^{\infty} dx_1 \, dx_2
\int \! \! \dots \! \! \int_{-\beta/2}^{\beta/2} d\tau_1 \! \ldots d\tau_4
\cf{\tilde{J}^z(\tau_1, x_1)
\tilde{J}^z(\tau_1, x_2) {\tilde {\cal O}}(\tau_3, 0) {\tilde {\cal
O}}(\tau_4, 0)}_{T, c} . \label{chi-imp}
\end{eqnarray}

Let us consider the non-Fermi liquid scenario with ${\cal O}$ given by
(\ref{LCBO}). The $U(1)$, $SU(2)_2$ and Ising
sectors are decoupled, and hence there are no dynamical correlations
between operators belonging to different sectors. As ${\cal
O}$ contains only the identity as $SU(2)_2$ factor, it follows that
\begin{equation}
\cf{\tilde{J}^z(\tau_1, x_1) \ldots \tilde{J}^z(\tau_m, x_m)
{\tilde {\cal O}}(\tau, 0)}_{T,c} = 0 \ ,
\end{equation}
using that the {\em LCBO} is Virasoro primary,
i.e. $\cf{ \tilde{\cal O}(w)}_{T} = 0$. The expectation value of
$\tilde{J}^z$ vanishes by the same reason, and we may decompose
\begin{eqnarray}
\cf{J^z_1 J^z_2 {\cal O}_3 {\cal O}_4}_c
& = & \cf{J^z_1 J^z_2 {\cal O}_3 {\cal O}_4}
- \cf{J^z_1 J^z_2} \cf{{\cal O}_3 {\cal O}_4} \nonumber \\
& - & \cf{J^z_1 {\cal O}_3} \cf{J^z_2 {\cal O}_4}
- \cf{J^z_1 {\cal O}_4} \cf{J^z_2 {\cal O}_3} = 0.
\label{decomp}
\end{eqnarray}
Hence, we infer that ${\cal O}$ does not give any contribution to
$\chi_{imp}$ to $O(\lambda_I^2)$. The same conclusion holds for {\em any}
candidate {\em LCBO} obtained in Sec.~\ref{section5C}, as all of them
are Virasoro primary and contain only the identity as $SU(2)_2$ factor.
Higher order terms in an expansion of
$\chi_{imp}$ in $\lambda_I$ can similarly be shown to vanish. The leading
contribution to $\chi_{imp}$ is instead given by the lowest-dimension
boundary operator that has a {\em non-trivial} $SU(2)_2$ factor. By our
symmetry analysis in Sec.~\ref{section5C}, this is given by
the $SU(2)_2$ energy-momentum tensor
$T_3(z) = \frac{1}{4} \no{\vJ(z) \cdot \vJ(z)}$,
of dimension $\Delta=2$. Analogous to the ordinary Kondo problem
\cite{Lrev}, $T_3(z)$ produces a finite impurity susceptibility (to first
order in $\lambda_I$),
\begin{equation}
\chi_{imp} \sim  T^0 + O(T). \label{lastequation}
\end{equation}
Thus, comparing with the free case (\ref{FLscaling}), the
electron-electron interaction is seen {\em not} to influence $\chi_{imp}$:
the impurity remains completely screened, in agreement with
the result of Furusaki and Nagaosa \cite{FN}.
Also note that, by (\ref{nr2}) and (\ref{lastequation}), the
favored non-Fermi liquid scenario implies a Wilson ratio that is
non-universal, and depends on temperature.

\section{Summary}
\label{section6}

We have studied the low-temperature properties of a spin-\onehalf\ magnetic
impurity coupled to a one-dimensional interacting electron system.
By turning the problem into a boundary critical phenomenon and using
conformal field theory we have reached the important conclusion
that the symmetry of the problem admits only one of two possible fixed points
describing the local electron-impurity composite: {\em Either} the
theory remains a local Fermi liquid in the presence of electron-electron
interaction (as for the ordinary Kondo problem with free electrons) {\em or}
electron correlations drive the system to a new fixed point with an anomalous
specific heat, identical to that proposed recently by Furusaki and
Nagaosa \cite{FN}. We have also shown that the suppression of
back scattering off the impurity destabilizes {\em both} fixed points and
produces an impurity critical behavior identical to that of the two-channel
Kondo model, but with a new Wilson ratio.

The non-Fermi liquid fixed point is distinguished by the presence of a
leading-correction-to-scaling operator of dimension
$\Delta = \Delta_c + \Delta_s = \onehalf e^{2\theta} + \onehalf $,
corresponding to the energy level $E'$, with\footnote{Note that we here
refer to the spectrum of an auxiliary problem with two impurities, one at
each end of the interval $[0,\ell]$. Cf. the discussion on {\em
double fusion} in the text after Eq. (\ref{selectionIII}) and (\ref{sr6}).}
\begin{equation}
E' - E_0 = \frac{\pi v_c}{ \ell}\onehalf e^{2\theta} + \frac{\pi
v_s}{ \ell}\onehalf. \label{signal}
\end{equation}
By numerically computing the finite-size energy spectrum of the
two-impurity auxiliary problem and checking
for the presence of $E'$, one has, in principle, a diagnostic
tool for deciding which scenario is realized: Fermi- or non-Fermi
liquid. Unfortunately, with our approach we cannot derive a unique
and complete finite-size spectrum at the non-Fermi liquid fixed
point as the selection rule in (\ref{sr4}) only applies to the
charge sector; there are still a multitude of ways of coupling the
two $U(1)$ conformal towers to those in the $SU(2)_2$ and Ising
sectors. When deriving possible $LCBO$ dimensions we were helped by
symmetry constraints, which, however, are no longer applicable when
considering the full spectrum. This fact makes the check against
numerics more difficult, as one is essentially restricted to search
for the single level $E'$.

To make contact with future experiments clearly requires a more complete
description of the system, as well as the inclusion of potential
scattering off the impurity (see Sec.~\ref{section1}). In the case of a
magnetic defect
implanted in a quantum wire, the observable of
greatest interest is the shift of the average conductance due to
the impurity, as this is the quantity most easily accessible in the
laboratory. Other important characteristics include
the local spin and charge Green's functions, the
scattering matrix, and the residual entropy. Considering the success of
conformal field theory techniques for obtaining these quantities in the
multi-channel Kondo
problem \cite{AL1}, we judge that the approach as presented in this
paper will be equally powerful. We hope to return to these, and related
issues in a future publication.

\section*{Acknowledgments}

It is a pleasure to thank A. A. Nersesyan and E. Wong for many inspiring
and illuminating discussions. We are also indebted to I. Affleck,
M. P. M. den Nijs, S. Eggert, T. Giamarchi, D. Kim, A. W. W. Ludwig,
and A. M. Tsvelik for useful comments and suggestions.
This research was supported by NSF grants DMR-9205125 and
DMR-91-120282 (P. F.), and a grant from the Swedish Natural
Science Research Council (H. J.).

\newpage


\end{document}